\documentclass[11pt]{article}
\usepackage{latexsym,amsmath,amscd,amssymb, graphics}
\usepackage{amsmath,amssymb,mathrsfs,framed,esint,slashed,color}
\usepackage{amsthm}
\usepackage{enumerate}
\usepackage[colorlinks]{hyperref}
\usepackage{enumitem}
\usepackage[margin=1in]{geometry}
\usepackage{stmaryrd}
\usepackage{graphicx}
\usepackage[all]{xy}
\usepackage[makeroom]{cancel}

\usepackage{framed}

\usepackage{pgfplots, pgfplotstable, booktabs, colortbl, array,multirow}
\usetikzlibrary{calc}
\usepgfplotslibrary{groupplots}
\pgfplotsset{compat=newest}

\theoremstyle{plain}
\newtheorem{theorem}{Theorem}[section]

\theoremstyle{definition}

\newtheorem{remark}{Remark}[section]

\usepackage{verbatim}

\begin{document}

\title{Systems, variational principles and interconnections in nonequilibrium thermodynamics}

\author{Fran\c{c}ois Gay-Balmaz\thanks{\noindent CNRS and \'Ecole Normale Sup\'erieure, Laboratoire de M\'et\'eorologie Dynamique, Paris, France, \href{francois.gay-balmaz@lmd.ens.fr}{francois.gay-balmaz@lmd.ens.fr}} \; and \; Hiroaki Yoshimura\thanks{\noindent School of Science and Engineering, Waseda University. 3-4-1, Okubo, Shinjuku, Tokyo, Japan, \href{yoshimura@waseda.jp}{yoshimura@waseda.jp}}}

\date{}

\maketitle

\begin{abstract}
The paper investigates a systematic approach to modeling in nonequilibrium thermodynamics by focusing upon the notion of interconnections, 
where we propose a novel Lagrangian variational formulation of such interconnected systems by extending the variational principle of Hamilton in mechanics.  In particular, we show how a nonequilibrium thermodynamic system can be regarded as an {\it interconnected system} of primitive physical elements or subsystems throughout an {\it interconnection}. While this approach is \textcolor{black}{new in nonequilibrium thermodynamics, this idea has been known as a useful tool for the modeling of complicated systems in networks as well as in mechanics. Hence, the} setting developed in this paper yields a promising direction for building a unifying description in various areas of modern science via thermodynamic principles, while being at the same time related to the early developments of variational mechanics.
\end{abstract}

\section{Introduction}

\subsection{Variational principle in thermodynamics}

One of the most fundamental principle in physics is the Principle of Critical Action. Maxwell's equations in electromagnetism, Newton's equations of motion in classical mechanics, Shr\"odinger's equations in quantum mechanics, Einstein's equations in general relativity, can all be obtained by extremizing a quantity, called action functional, encoding all the properties of the system. For \textit{classical mechanics}, this principle reduces to Hamilton's principle, which enables us to systematically formulate the dynamics of conservative systems, \cite{Lanczos1986}. For the cases in which a mechanical system involves a nonconservative external force and kinematic constraints, whether they are holonomic or not, the variational principle of Hamilton is replaced by the Lagrange-d'Alembert Principle, where the additional term associated to the external force must be included and in which the variations of the curves are subject to the so-called variational constraints.
Regarding \textit{nonequilibrium thermodynamics}, on the other hand, which includes mechanics and is also deeply related to various disciplines such as computational fluid dynamics, chemical reactions, biological and engineering systems,\cite{deGrootMazur1969,KoPr1998}, there has been a large gap with mechanics from the viewpoint of the proposed variational formulations. In other words, a variational formulation for nonequilibrium thermodynamics that includes the Hamilton action principle in mechanics as a special case has not been completely established.
In fact, although there have been proposed many variational approaches such as the \textit{minimum dissipation principle} by \cite{Onsager1931, OnMa1953, MaOn1953} and the \textit{minimum entropy production principle} by \cite{Prigogine1947, GlPr1971}, these principles mainly treat the entropy production and the related dissipative energy of the thermodynamic system, while one does not know how to recover Hamilton's principle in mechanics from these approaches when the irreversible processes are absent.
In particular, it has not been clarified how to link Hamilton's variational principle in mechanics with those variational approaches in thermodynamics. Another relevant work was done by \cite{Bi1975, Bi1984}, in which a generalized form of d'Alembert principle called a {\it principle of virtual dissipation} was illustrated with applications to irreversible thermodynamic systems with viscoelasticity, thermoelasticity as well as heat transfer. However, it was limited to {\it weakly irreversible systems}, {\it isothermal systems or quasi-isothermal systems}, where the required constraints associated with the rate of entropy production were simplified to be holonomic or quasi-holonomic. \textcolor{black}{There exists some other variational formulation in thermodynamics including mesoscopic views, including Boltzman kinetic equations, see \cite{Grm2014}. However, it is not an extension of the variational principle of Hamilton in Lagrangian mechanics.} 
Due to the strong connection of nonequilibrium thermodynamics with various disciplines, the understanding of this field from a variational formulation closely related to that of mechanics would also foster the identification of common principles between different disciplines and provide a promising tool for the modelling of multiphysical systems. The questions emerging from the above comments can be summarized as follows:\vspace{-0.2cm}
\begin{itemize}
\item What is the variational principle for nonequilibrium thermodynamics, consistent with Hamilton's principle in mechanics?
\item How can we systematically understand interconnection in the realm of nonequilibrium thermodynamic systems via variational principles?
\end{itemize}\vspace{-0.2cm}
To answer these questions, we will first show how the entropy production associated to irreversible processes can be understood as a {\it phenomenological constraint} that is nonlinear nonholonomic and also how it is incorporated into a variational principle of the Lagrange-d'Alembert type, see \cite{GBYo2017a,GBYo2017b,GBYo2019b}. Second, with the help of {\it network theory}\cite{Kr1963,OsPe1973,OsPeKa1973}, we will introduce the notion of {\it interconnections} in order to represent the structure of nonequilibrium thermodynamic systems, by which one can reticulate the original system into an {\it interconnected system} of primitive systems or constituent elements throughout the energy flow. Mathematically the interconnection is represented by a Dirac structure, which is an extended notion of symplectic and Poisson structures, and provides us with an implicit relation between dual variables, namely, velocity (vector) and force (covector). It has been clarified that the constraints due to Kirchhoff Current Law and Kirchhoff Voltage Law in electric circuit and those due to nonholonomic constraints can be modeled by such interconnections, see \cite{YoMa2006a}. However, it has not been \textcolor{black}{systematically} understood how such interconnections appear in thermodynamics. In this paper, \textcolor{black}{we propose a novel idea of how} the interconnection can incorporate the nonlinear phenomenological relations linked to the entropy production. It is then showed that the variational formulation of a  nonequilibrium thermodynamic system can be systematically constructed by the interconnection of the variational formulation of each  primitive system.

The approach that we develop in this paper is relevant for the modelling and analysis of complex systems. On one hand, the analysis of the interconnection helps recognizing the primitive entities, denoted $ \boldsymbol{\Sigma}_k$, $k=1,...,P$, which can be of different nature (mechanical, thermal, electrical), and whose dynamics is easier to design and understand than the original system $\boldsymbol{\Sigma}=\cup_{k=1}^P\boldsymbol{\Sigma}_k$. On the other hand it allows to correctly describe the known and unknown energetic fluxes among such primitive entities $\boldsymbol{\Sigma}_k$ and $\boldsymbol{\Sigma}_\ell$, which must be consistent with the fundamental laws of thermodynamics. It has been argued that such an interconnection approach may be potentially useful beyond the setting of physical science by inspiring modeling techniques for social sciences, such as the dynamics of human society evolution, where the concepts of energy, power, and fluxes may be given some interpretations, see \cite{PoMG2019} and references therein. The development of a variational formulation for primitive systems in thermodynamics and the systematic understanding of the interconnection of such variational formulations to achieve a description of the full system thus provides a promising tool for building a common description in various areas of modern science. At the same time, such variational formulations have their roots in the critical action principle of Hamilton, and are thus connected to the early developments of classical mechanics.

The variational description also naturally yields the associated \textit{geometric structures} underlying the dynamics of each subsystems as well as the interconnections of such subsystems. While we shall not consider this aspect in this paper, let us mention that a special class of Dirac structures that is induced from the interconnection constraints, called an {\it interaction Dirac structure}, behaves well in interconnecting subsystems in the sense that the variational structures associated to each subsystem are to be intertwined through the interaction Dirac structures $D_{\rm int }$ associated to the interconnection conditions. This naturally produces a variational formulation of the interconnected system which yield the equations of motion for the overall system.
 
\section{System theoretic approach to thermodynamics}
\subsection{Nonequilibrium thermodynamic systems}
\begin{figure}[h]
\vspace{-0.3cm}
\begin{center}
\includegraphics[scale=.4]{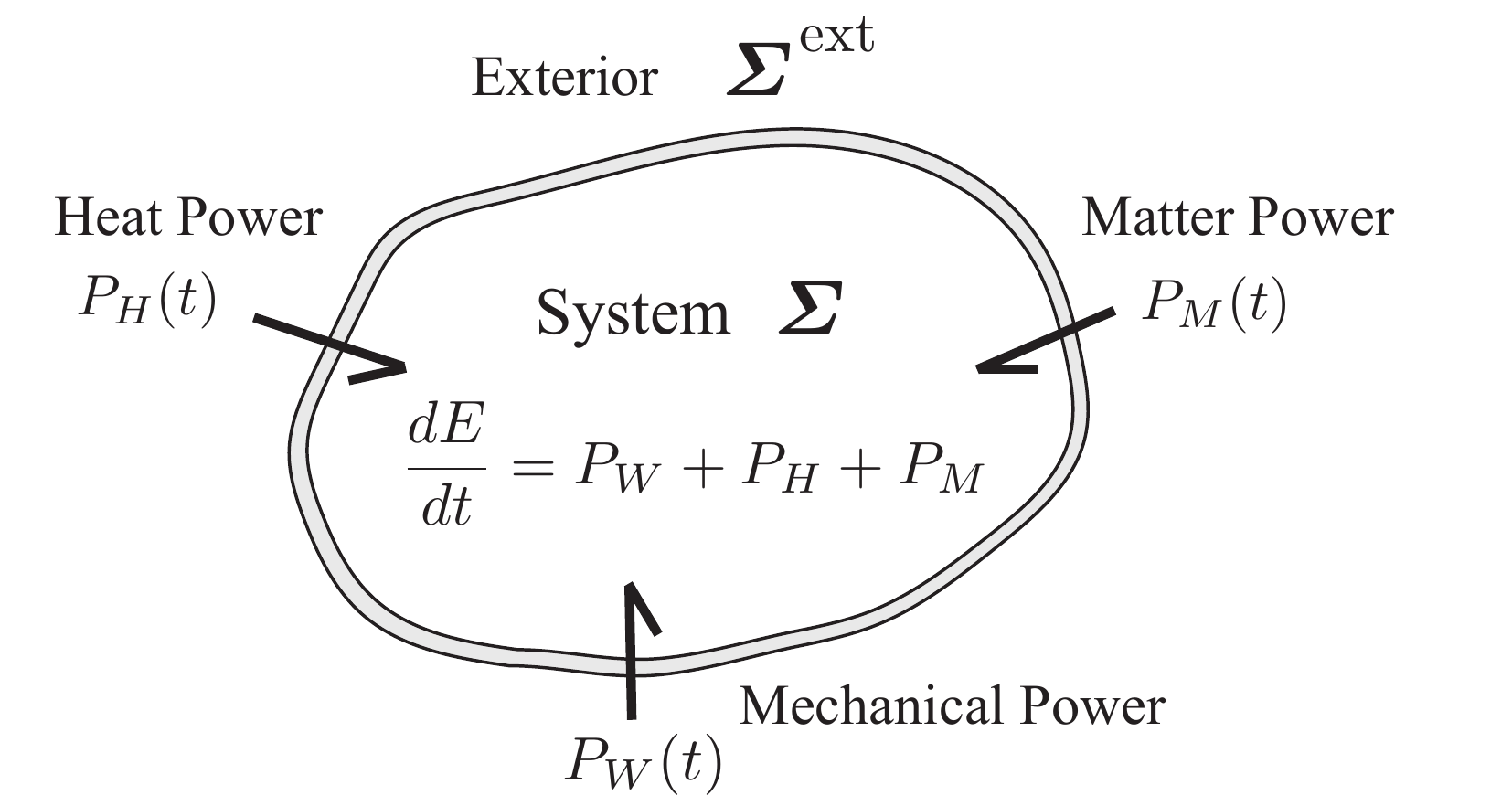} 
\caption{System with energetic interactions} \label{fig_System}
\end{center}
\vspace{-1cm}
\end{figure}

Let us consider a thermodynamic system $ \boldsymbol{\Sigma} $ which has energetic interactions with its exterior $ \boldsymbol{\Sigma} ^{\rm ext}$, as shown in Figure \ref{fig_System}. The state of the thermodynamic system is described by a set of mechanical state variables $(\mathbf{q},\mathbf{v})$ and thermodynamic variables $\zeta$. Functions of these variables are referred to as state functions. Let us call $ \boldsymbol{\Sigma}  $ a \textit{simple} system if the thermodynamic state is described in terms of a single variable $\zeta$, usually, chosen by an \textit{entropy} $S$ as will be shown. Let $P^{\rm ext}_M(t)$ be the power exchange with exterior which is associated to the matter transfer, $P^{\rm ext}_H(t)$ that associated to the heat transfer, and  $P^{\rm ext}_M(t)$ that associated to the mechanical force.

Any thermodynamic system can be classified into the following cases:
\begin{itemize}
\item $ \boldsymbol{\Sigma} $ is {\bf closed} if $P^{\rm ext}_M(t)=0$. \vspace{1mm}
\item 
$ \boldsymbol{\Sigma} $ is  {\bf adiabatically closed} if $P^{\rm ext}_M(t)=P^{\rm ext}_H(t)=0$. \vspace{1mm}
\item 
$ \boldsymbol{\Sigma} $ is  {\bf  isolated} if $P^{\rm ext}_M(t)=P^{\rm ext}_H(t)=P^{\rm ext}_W(t)=0$.
\end{itemize}

\subsection{Laws of nonequilibrium thermodynamics}
\paragraph{First law.}
Following \cite{StSc1974}, the first law states that for every system $ \boldsymbol{\Sigma} $, there exists an extensive state function $E$, which satisfies the following relation
\begin{equation}\label{1st_law} 
\frac{d}{dt} E(t) = P^{\rm ext}_W(t)+P^{\rm ext}_H(t)+P^{\rm ext}_M(t),
\end{equation} 
where $t$ denotes {\it time}. This function is called the \textit{energy} of the system. In the case of {\it an isolated system}, the energy is conserved.
\paragraph{Second law.}
For every system $ \boldsymbol{\Sigma} $ there exists an extensive state function $S$, which satisfies the following relations
\begin{itemize}
\item[(i)]  
If $ \boldsymbol{\Sigma} $ is adiabatically closed, $S$ is a non-deceasing function with respect to $t$, i.e., 
\[
\frac{d}{dt} S(t)=I(t)\geq 0,
\]
where $I(t)$ is the {\it internal entropy production rate} linked to the irreversible processes.
\item[(ii)] 
If $ \boldsymbol{\Sigma}$ is isolated, $S$ tends to a finite local maximum as $t$ goes to infinity, namely,
\[
\lim_{t \rightarrow +\infty}S(t)= \max_{ \rho \; \text{compatible}}S[\rho ],
\]
where ``$ \rho $ compatible" indicates a thermodynamic state that is compatible with the constraints, such as isolation conditions and internal walls.
\end{itemize}
This function is called the \textit{entropy} of the system.

For the case in which $ \boldsymbol{\Sigma}$ is isolated, it is said to be {\it reversible} if $I(t)=0$, i.e., $S$ is constant. For the case in which the system is not isolated, it is said to be {\it reversible} if 
the total isolated system (formed by the system and the surrounding with which it interacts) is reversible.

\subsection{Systems and interconnection}

By definition, a \textit{system} is a set of \textit{objects} with \textit{relationships} between the objects and between their \textit{attributes}, see \cite{HaFa1956, Paynter1961}. In any physical system, the relationships are given in terms of power flow, i.e., throughout the energy balance, and we can interpret the objects as constituent elements of the system such as mass, spring and friction for mechanical systems, or inductors, capacitors and resistors in electric circuits, etc. Their attributes may be interpreted by the constitutive relations of physical components. The structural relations between the objects, namely, elements or subsystems, can be modeled as an \textit{interconnection}, whose relationships are given by \textit{input-output relation} between \textit{dual variables} $\mathbf{v}$ and $\mathbf{f}$, such as \textit{velocity} and \textit{force} variables in mechanics, \textit{current} and \textit{voltage} variables  in circuits, or \textit{entropy flow} and \textit{temperature} in thermodynamics. The associated power flow $P$ given by the dual paring as $P=\left<\mathbf{f}, \mathbf{v}\right>$ vanishes, which describes the power invariance or energy balance. An instance of this relation is Tellegen's theorem in circuit theory, see \cite{DeKu1969}.

Such structural relations can be mathematically understood by symplectic, Poisson or Dirac structures on the state space $M$ and they are given among a vector field (velocity field) $X$ and a one-form (force field) $\alpha$ on $M$, where $M$ is usually the phase space, or the cotangent bundle over a configuration manifold. For instance, for a symplectic structure $\Omega: TM \times TM \to \mathbb{R}$, the structural relation is given by $\alpha=\Omega^{\flat} X$ where $\Omega^{\flat}: TM \to T^{\ast}M$ is given by $\langle\Omega^{\flat}(z)(X_z), Y_z\rangle =\Omega(z)(X_z, Y_z)$, $ z \in M$, $ X _z, Y _z \in T_zM$. For a Poisson structure $B: T^*M \times T^*M \to \mathbb{R}$, the structural relation is given by $X=B^{\sharp} \alpha$, where the associated bundle map $B^{\sharp}: T^{\ast}M \to TM$ is defined by $B(z)(\alpha _z,  \beta _z)= \langle \alpha _z, B^{\sharp}(\beta _z) \rangle$, $ z \in M$, $ \alpha _z, \beta _z \in T^*_zM$. Finally, for a Dirac structure $D \subset TM\oplus T^\ast M$, the structural relations are implicitly given as $(X, \alpha) \in D$. In particular, such a structure is called an {\it interconnection} among elements or subsystems, each of which is called a {\it primitive system}; see \cite{Kr1963,YoMa2006a,JaYo2014}. \textcolor{black}{In Hamiltonian mechanics, for example, the force field $\alpha$ is usually given by the differential of a given Hamiltonian function $H$ on $M$ and hence the dynamics is given in the context of symplectic structures as $\mathbf{d}H(z)=\Omega^{\flat}(z)\cdot X_H(z)$ for each $z \in M$, while it is given as $X_H(z)=B^{\sharp}(z) \mathbf{d}H(z)$ in the framework of Poisson structures. Notice that $B^{\sharp} =(\Omega^{\flat})^{-1}$ on a symplectic manifold $M$, which implies that the input and output relations are given explicitly between the force field $\mathbf{d}H$ and the velocity field $X_H$ but they are given in converse for symplectic and Poisson structures. For those cases, the Dirac structure may be given by $D=\mathrm{graph}\,\Omega^{\flat}$ or $D=\mathrm{graph}\,B^{\sharp}$, where the input and output relations are implicitly given between $\mathbf{d}H$ and $X_H$. On symplectic manifolds, by using the canonical coordinates $q^i, p_i,\,i=1,...,n$ for $M$, which is ensured by Darboux's theorem, we can recover the usual Hamilton equations as $\dot{q}^i= \frac{\partial H}{\partial p_i}$ and $\dot{p}_i= -\frac{\partial H}{\partial {q}^i}$.}

%\todo{FGB: I think I don't understand the blue sentence above "and it means the power invariance or the energy balance". }

%\todo{\textcolor{red}{HY: $P=\left<\mathbf{f}, \mathbf{v}\right>=\sum_{i=1}^{N}f_i v^i=0$ denotes {\it Tellegen's theorem} in circuit theory, which is known as the power invariance (which is also true in mechanics) associated to the connection $n$-port  or non-energic $n$-port, see some textbook such as "Basic Circuit Theory by Desoer" or the paper "A theory of nonenergic $N$-ports" by Wyatt and Chua(1975);  Oster and Perelson(1973). See also YoMa2006a. This implies that the pysical object of a Dirac structure is a non-energic $n$-port or connection $n$-port in circuits.  In systems theory, more generally, we call it an interconnection. For instance, in the case of conservative systems, $\frac{dE}{dt}=P=\left<F, \dot{q}\right>=\sum_{i=1}^{N}F_i \dot{q}^i=0$, where $F_i= \frac{d}{dt}\left( \frac{\partial L}{\partial \dot{q}^i}\right)- \frac{\partial L}{\partial q^i}$. For the case of non-conservative systems, see the situation in Fig.2, where $P=\left<F, \dot{q}\right>=\sum_{i=1}^{N}F_i \dot{q}^i=0$, while $P=(P_L+P_C)+(-P_V-PR)=0$. It means the energy balance equation $\frac{dE}{dt}=P_L+P_C=P_V+P_R$.}}

\begin{figure}[h]
\vspace{-0.3cm}
\begin{center}
\includegraphics[scale=.5]{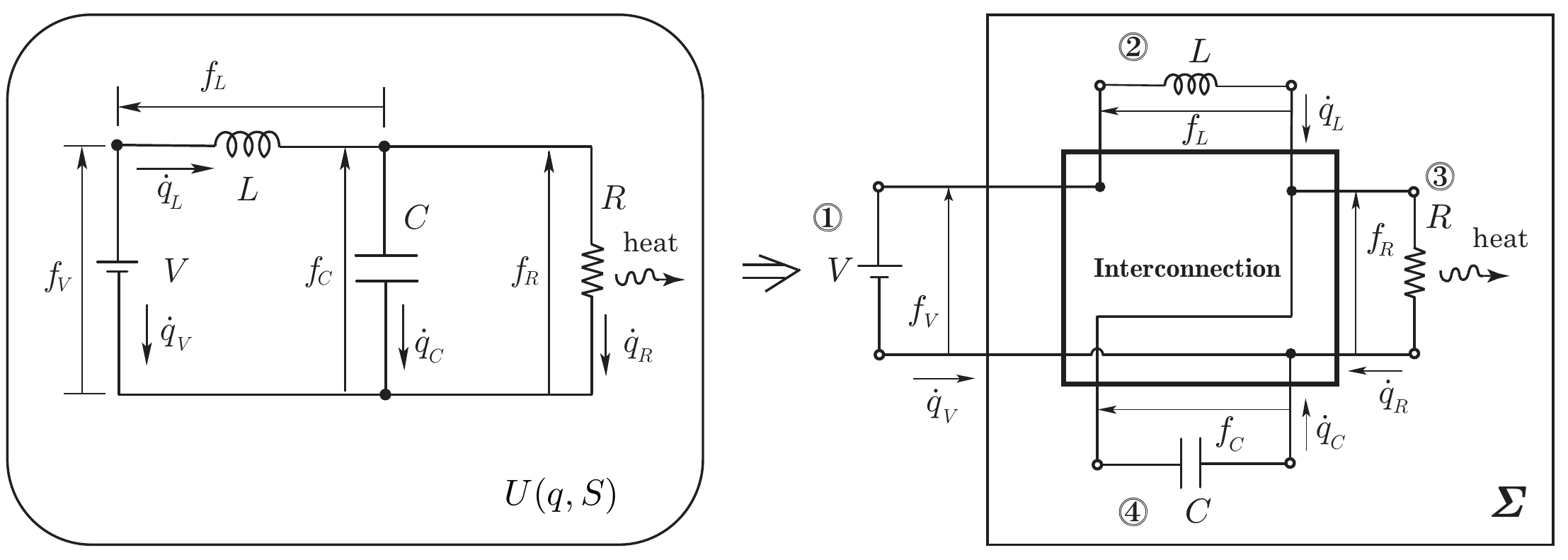} 
\caption{Interconnection of an L-C-R circuit} \label{fig_INT_LCR}
\end{center}
\vspace{-0.5cm}
\end{figure}

\paragraph{Example.}
As an illustration, consider the thermodynamics of an  electric circuit which consists of an inductor L, a capacitor C, a resistor R and a source of voltage V, as shown on Figure \ref{fig_INT_LCR}. We include the internal entropy production regarding the irreversibility of the resistor. The configuration space is $Q\times \mathbb{R}  =\mathbb{R}^5$ which contains the charges $\mathbf{q}=(q^{1},q^{2},q^{3},q^{4})=(q_V,q_L,q_R,q_C) \in Q =\mathbb{R}^4$ as well as the entropy $S \in \mathbb{R} $.

The circuit can be understood as an interconnected system throughout the interconnection $\Delta_Q \times \Delta_Q^\circ \subset TQ \oplus T^\ast Q$, where the distribution $\Delta_Q
 \subset TQ\footnote{We recall that a distribution $ \Delta _Q$ on a manifold $Q$ is a vector subbundle of $TQ$. Roughly speaking, a distribution assigns to each point $ \mathbf{q}\in Q $ a vector subspace $ \Delta _Q( \mathbf{q} ) \subset T_ \mathbf{q} Q$ in a smooth way, where it is assumed that each subspace at $\mathbf{q}$ has the same dimension. In the present case $Q= \mathbb{R} ^4$ is a vector space and the distribution does not depend on $ \mathbf{q} $, so it is just a vector subspace of $ \mathbb{R} ^4$.}$ describes the Kirchhoff circuit law (KCL) constraints among currents $\mathbf{\dot{q}}=(\dot{q}^{1},\dot{q}^{2},\dot{q}^{3},\dot{q}^{4})=(\dot{q}_V,\dot{q}_L,\dot{q}_R,\dot{q}_C) \in T_ \mathbf{q} Q$, and where its annihilator $\Delta_Q^{\circ}$ defined by $ \Delta _Q^\circ( \mathbf{q} ) :=\{\mathbf{f} \in T^*_ \mathbf{q} Q\mid \left< \mathbf{f}, \mathbf{\dot{q}} \right>=0,\;\;\forall \,\mathbf{\dot{q}} \in \Delta_Q( \mathbf{q} )\}$
describes the Kirchhoff voltage law (KVL) constraints among voltages $\mathbf{f}=(f_{1},f_{2},f_{3},f_{4})=(f_V,f_L,f_R,f_C) \in T_ \mathbf{q} ^{\ast}Q$. For the present case, see Figure \ref{fig_INT_LCR}, one has
\begin{equation}\label{KCL_const}
\Delta _Q( \mathbf{q} ) = \{  \mathbf{\dot{q}}  \in T_ \mathbf{q} Q \mid  \langle   \omega^{a}, \mathbf{\dot{q}}  \rangle=0, a=1, 2\},
\end{equation}
with $\omega^{a}=\omega_{k}^{a} \, dq^{k}, \;a=1,2; \; k=1,...,4,$ the two independent covectors given in matrix representation by
\[
\omega_{k}^{a}=
\left(
\begin{array}{cccc}
-1 & -1 & 0 & 0 \\
0 & 1 & -1 & -1 \\
\end{array}
\right).
\]
The interconnection $D= \Delta _Q \times \Delta _Q^\circ  \subset TQ \oplus T^\ast Q$ is an example of an interaction Dirac structure, see \cite{JaYo2014}.

Furthermore, there exists an internal entropy production associated to the irreversibility of the resistor, i.e., 
$$
I=\dot{S}=-\frac{1}{T}V_R(\mathbf{q}, \dot{\mathbf{q}}, S) \cdot \dot{\mathbf{q}},
$$
where $\dot{(\;)}$ indicates the derivative with respect to time,  $T$ is the temperature determined from the internal energy $U=U(\mathbf{q},S)$ of $\boldsymbol{\Sigma}$ as $T= \frac{\partial U}{\partial S}$, and the constitutive relation of Resistor R is given by Ohm's law as $V_R(\mathbf{q}, \dot{\mathbf{q}}, S)=-R(\mathbf{q},S)\cdot \dot{\mathbf{q}}$. Here $R(\mathbf{q},S)$ is a positive coefficient of the resistance, so that $I\ge0$ follows for each time $t$.
We shall treat this example in the context of variational formulations later.

\section{Variational formulation of thermodynamic systems}\label{Sec_VF_thermodynamics}
The variational formulation of thermodynamics that we consider here is an extension of the critical action principle in mechanics, namely, Hamilton's principle. We review below Hamilton's principle and its modifications to handle forced and constrained holonomic or nonholonomic \textit{mechanical} systems, before considering the case of simple and non-simple \textit{thermodynamic} systems.

\subsection{Variational principles in mechanics}
\paragraph{Hamilton's principle.} Consider a mechanical system with an $n$ dimensional configuration manifold $Q$, let $q^1,...,q^n$ be the local coordinates for $\mathbf{q} \in Q$, and consider a Lagrangian function $L:TQ \rightarrow \mathbb{R}$ defined on the velocity phase space or the tangent bundle $TQ$ of the manifold $Q$. Hamilton's principle states the critical condition for the action functional associated to $L$, namely: 
\[
\delta \int_{t_1}^{t_2} L( \mathbf{q}, \dot {\mathbf{q}} ) {\rm d} t =0,
\]
which must be satisfied for any variations $ \delta \mathbf{q}$ with $ \delta \mathbf{q} (t_1)= \delta \mathbf{q} (t_2)=0$. The critical curve $\mathbf{q}(t),\,t \in [t_1,t_2]$ solves the equations of motion for the mechanical system, given by the Euler-Lagrange equations
\begin{equation}\label{EuLa_eq} 
\frac{d}{dt} \frac{\partial L}{\partial \dot{\mathbf{q}}}- \frac{\partial L}{\partial \mathbf{q}}=0.
\end{equation}
\paragraph{Lagrange-d'Alembert principle for mechanical system with forces.} Hamilton's principle can be modified to handle the case in which a nonconservative external force field $ \mathbf{F} ^{\rm ext}:TQ \to T^\ast Q$ acts on the system, where $T^*Q$ denotes the phase space (or cotangent bundle) of the manifold $Q$. This results in the critical action principle
\begin{equation}\label{HP_forced} 
\delta \int_{t_1}^{t_2} L(\mathbf{q} , \dot {\mathbf{q}} ) {\rm d} t +\int_{t_1}^{t_2}\left\langle \mathbf{F} ^{\rm ext}(\mathbf{q}, \dot {\mathbf{q}} ),\delta{\mathbf{q}} \right\rangle  {\rm d} t =0,
\end{equation} 
for any $ \delta \mathbf{q}$ with $ \delta \mathbf{q} (t_1)= \delta \mathbf{q} (t_2)=0$. Here $ \left\langle \cdot , \cdot \right\rangle $ denotes the pairing between a covector in $T^*_ \mathbf{q} Q$ and a vector in $ T_ \mathbf{q} Q$ for $ \mathbf{q} \in Q$. The critical curve $\mathbf{q}(t),\,t \in [t_1,t_2]$ satisfies the Euler-Lagrange equations with the external force:
\begin{equation}\label{EuLa_eq_Force} 
\frac{d}{dt} \frac{\partial L}{\partial \dot{\mathbf{q}}}- \frac{\partial L}{\partial \mathbf{q}}=\mathbf{F} ^{\rm ext}.
\end{equation} 
\paragraph{Hamilton's principle for mechanical systems with holonomic constraints.}
Assume that the motion $\mathbf{q}(t)$ is restricted to an $n-m$ dimensional submanifold $N \subset Q$ given by $m$-functions $\Phi^r: Q \to \mathbb{R},\,r=1,...,m<n$ as $N=\{ \mathbf{q} \in Q \mid \Phi^r(\mathbf{q})=0, \,r=1,...,m<n\}$. In this case, the action functional in Hamilton's principle can be modified by adding the constraint, which results in the critical action principle
\[
\delta  \int_{t_1}^{t_2} \Big[ L( \mathbf{q} , \dot {\mathbf{q}} ) + \sum_{r=1}^m \lambda_r \Phi^r (\mathbf{q})\Big]  {\rm d} t=0,
\]
for arbitrary $ \delta \lambda _r$ and $ \delta \mathbf{q} $ with $ \delta \mathbf{q} (t_1)= \delta \mathbf{q} (t_2)=0$, where $\lambda_r, \, r=1,...,m$ are Lagrange multipliers. The critical curve $\mathbf{q}(t)$ then satisfies
\begin{equation}\label{EuLa_eq} 
\frac{d}{dt} \frac{\partial L}{\partial \dot {\mathbf{q}} }= \frac{\partial L}{\partial \mathbf{q}} +\sum_{r=1}^m
\lambda_r \frac{\partial \Phi^r}{\partial  \mathbf{q} },\qquad \Phi^r(\mathbf{q} )=0,\;\; r=1,...,m.
\end{equation} 
\paragraph{Lagrange-d'Alembert principle for systems with nonholonomic constraints.} Consider a mechanical system subject to a linear nonholonomic constraint given by a distribution $\Delta _Q$ on $Q$. The evolution $ \mathbf{q} (t) \in Q$ of the system must satisfy $ \dot{\mathbf{q}}(t) \in \Delta _Q( \mathbf{q} (t))$ for all $ t \in [t_1,t_2]$, typical examples are provided by rolling constraints; see \cite{Bl2003}.
%Consider a distribution on $Q$ given as $ \Delta_{Q}=\{\dot{\mathbf{q}} \in T_{\mathbf{q}}Q \mid \phi^{a}(\mathbf{q},\dot{\mathbf{q}})=\left< \omega^{a}(\mathbf{q}), \dot{\mathbf{q}}\right>=0\} \subset TQ$, where $\omega^{a}: Q \to T^{\ast}Q, a=1,...,r <n$ are given one-forms on $Q$.
To get the evolution equations for such systems, we consider the Lagrange-d'Alembert principle
\begin{equation}\label{LDA_VC} 
\delta \int_{t_0}^{t_1} L(\mathbf{q}, \dot {\mathbf{q}} ) {\rm d} t+\int_{t_1}^{t_2} \left< \mathbf{F} ^{\rm ext}(\mathbf{q}, \dot{\mathbf{q}}), \delta \mathbf{q} \right> {\rm d} t =0,
\end{equation}
for $ \delta \mathbf{q}$ subject to the condition $\delta{\mathbf{q}}(t) \in \Delta_Q( \mathbf{q} (t))$, with $ \delta \mathbf{q} (t_1)= \delta \mathbf{q} (t_2)=0$, where the critical curve $ \mathbf{q} (t),\,t \in [t_0,t_1]$ satisfies the kinematic constraint $\dot{\mathbf{q}}(t) \in \Delta_Q( \mathbf{q} (t))$. This principle yields the Lagrange-d'Alembert equations for the curve $ \mathbf{q} (t)$ as
\begin{equation}\label{LDAEq} 
\frac{d}{dt} \frac{\partial L}{\partial \dot{\mathbf{q}}}- \frac{\partial L}{\partial \mathbf{q}} - \mathbf{F} ^{\rm ext}\in \Delta_Q^\circ(\mathbf{q}), \qquad \dot{\mathbf{q}} \in \Delta _Q( \mathbf{q} ).
\end{equation} 

It is important to note that the variational formulation \eqref{LDA_VC} is not an usual critical action principle since it involves constraints on the allowed variations, namely, $ \delta \mathbf{q}$ must belong to the subspace $ \Delta _Q( \mathbf{q} )$ of $T_ {\mathbf{q}}Q$. The distribution $ \Delta _Q$ thus plays both the role of a \textit{variational constraint} on $ \delta \mathbf{q} $ and a \textit{kinematic constraint} on $ \dot {\mathbf{q}}$. As shown later, a similar setting underlies the variational description of thermodynamic systems.

\paragraph{Energy balance.}
Define the total energy function $E:TQ\times\mathbb{R}\rightarrow\mathbb{R}$ for the Lagrangian $L(\mathbf{q}, \dot {\mathbf{q}} )$ as
\begin{equation}\label{def_E_tot}
E(\mathbf{q},\dot{\mathbf{q}})= \left<\frac{\partial L}{\partial \dot{\mathbf{q}}},  \dot{\mathbf{q}}\right>- L(\mathbf{q}, \dot{\mathbf{q}}).
\end{equation}
We get
\[
\frac{d}{dt} E = \left\langle \frac{d}{dt}\frac{\partial L}{\partial \dot{\mathbf{q}}}- \frac{\partial L}{\partial \mathbf{q}}, \dot{ \mathbf{q} } \right\rangle= \left\langle \mathbf{F} ^{\rm ext}( \mathbf{q}, \dot{ \mathbf{q}}),  \dot{ \mathbf{q} } \right\rangle = P^{\rm ext}_W,
\]
along a solution $ \mathbf{q} (t)$ of \eqref{LDAEq}, where $P_W^{\rm ext}$ is the power coming from the work done on the system by the external force $\mathbf{F} ^{\rm ext}( \mathbf{q}, \dot{ \mathbf{q}})$. This is the statement of the first law \eqref{1st_law} for that system.

\subsection{Variational formulation of simple thermodynamic systems}

Consider a \textit{simple} thermodynamic system, in which we recall that the state of the system can be described by a single entropy $S \in \mathbb{R}$ together with the mechanical state variables $(\mathbf{q},\dot{\mathbf{q}}) \in TQ$. Assume  that the mechanical motion is constrained by a given distribution $ \Delta _Q \subset TQ$ in the sense that a solution curve must satisfy $\dot{\mathbf{q}}(t) \in \Delta_Q(\mathbf{q}(t)) \subset T_{\mathbf{q}(t)}Q$. For such simple systems, the Lagrangian is a function
\[
L: TQ \times \mathbb{R}  \rightarrow \mathbb{R} , \quad (\mathbf{q}, \dot{\mathbf{q}}, S) \mapsto L(\mathbf{q}, \dot{\mathbf{q}}, S).
\]
Suppose that there exist external and friction forces $\mathbf{F} ^{\rm ext}, \mathbf{F} ^{\rm fr}:TQ\times \mathbb{R} \rightarrow T^* Q$. The variational formulation for the simple system is given by the following theorem, see also \cite{GBYo2017a}.

\begin{theorem}[Variational formulation for simple thermodynamic systems]\label{thm_simple_thermodynamic}
The following statements are equivalent:
\begin{itemize}
\item[(i)] The curves $\mathbf{q}(t)$, $S(t)$ are critical for the \textit{action functional}
\begin{equation}\label{LdA_thermo_simple} 
\delta \int_{t _1 }^{ t _2}L(\mathbf{q} , \dot{\mathbf{q}} , S){\rm d}t +\int_{t_1}^{t_2} \left< \mathbf{F} ^{\rm ext}(\mathbf{q}, \dot{\mathbf{q}}, S), \delta \mathbf{q} \right>\,dt =0,
\end{equation}
subject to the following \textit{kinematic constraints}
\begin{equation}\label{CK_simple} 
\dot{\mathbf{q}} \in \Delta_Q(\mathbf{q}), \qquad \frac{\partial L}{\partial S}(\mathbf{q}, \dot{\mathbf{q}}, S)\dot S  =  \left< \mathbf{F} ^{\rm fr}(\mathbf{q}, \dot{\mathbf{q}}, S), \dot{\mathbf{q}} \right>,\qquad 
\end{equation}
and for variations subject to the following \textit{variational constraints}
\begin{equation}\label{CV_simple} 
\delta{\mathbf{q}} \in \Delta_Q(\mathbf{q}), \qquad \frac{\partial L}{\partial S}(\mathbf{q}, \dot {\mathbf{q}}, S)\delta S=  \left< \mathbf{F} ^{\rm fr}(\mathbf{q} , \dot{\mathbf{q}} , S), \delta \mathbf{q} \right>,
\end{equation}
with $ \delta \mathbf{q}(t_1)=\delta \mathbf{q}(t_2)=0$.
\medskip 

\item[(ii)] The curves $\mathbf{q}(t)$, $S(t)$ solve the system of equations
\begin{equation}\label{simple_systems} 
\left\{
\begin{array}{l}
\displaystyle\vspace{0.2cm}\frac{d}{dt}\frac{\partial L}{\partial \dot{\mathbf{q}}}- \frac{\partial L}{\partial \mathbf{q}} - \mathbf{F} ^{\rm fr}(\mathbf{q}, \dot{\mathbf{q}}, S)- \mathbf{F} ^{\rm ext}(\mathbf{q}, \dot{\mathbf{q}}, S) \in \Delta_Q^{\circ}(\mathbf{q}),\\
\displaystyle \dot{\mathbf{q}} \in \Delta_Q(\mathbf{q}), \qquad \frac{\partial L}{\partial S}\dot S=  \left< \mathbf{F} ^{\rm fr}(\mathbf{q}, \dot{\mathbf{q}}, S), \dot{\mathbf{q}} \right>.
\end{array} \right.
\end{equation} 
\end{itemize}
\end{theorem}
This variational formulation includes the variational principle of Hamilton in mechanics as a particular case since the irreversible processes are incorporated into the Lagrange-d'Alembert equations that involve external and friction forces. For the case in which the entropy variable is not included, the variational formulation \eqref{LdA_thermo_simple}--\eqref{CV_simple} clearly reduces to \eqref{LDA_VC}. When the nonholonomic mechanical constraint $ \Delta _Q$ is absent, then it further restricts to \eqref{HP_forced}, and to the Hamilton principle itself when $ \mathbf{F} ^{\rm ext}=0$.

The temperature is given by the minus derivative of $L$ as to $S$, namely, $T=-\frac{\partial L}{\partial S}$, which must be always positive. When the Lagrangian is given by the kinetic energy $K( \mathbf{q} , \dot{ \mathbf{q} })$ minus the internal energy $U(\mathbf{q},S)$, namely, $L(\mathbf{q},\dot{\mathbf{q}}, S)= K(\mathbf{q}, \dot{\mathbf{q}}) - U(\mathbf{q},S)$, we can recover the standard definition of the temperature in thermodynamics as $T=-\frac{\partial L}{\partial S}=\frac{\partial U}{\partial S}$.
 
If the friction force is absent, it follows from the third equation in \eqref{simple_systems} that the entropy is to be constant $S_0$. Hence the system \eqref{simple_systems} becomes the forced Lagrange-d'Alembert or Euler-Lagrange equations in mechanics, see \eqref{LDAEq} or \eqref{EuLa_eq_Force}, for a Lagrangian that parametrically depends on $S_0$.

Finally, we note that the variational structure here is similar to the structure of the Lagrange-d'Alembert principle in nonholonomic mechanics because there are two kinds of constraints, i.e., the kinematic constraint \eqref{CK_simple} on the critical curve and the variational constraint \eqref{CV_simple} for the variations of curves. As in the Lagrange-d'Alembert case, one formally passes from the variational to the kinematic constraint by replacing $ \delta $-variations by time rate of change, such as $ \delta \mathbf{q} \rightarrow \dot  { \mathbf{q} }$ and $ \delta S \rightarrow \dot  S$. More strictly speaking, the nonholonomic constraints associated with the internal entropy production fall into the category of nonlinear constraints of thermodynamic type and refer to \cite{GBYo2017a} about the variational structures in details. This constraint involves the friction force, of phenomenological nature, and is hence referred to as a \textit{phenomenological constraint}.

\paragraph{Energy balance.} In a similar way with the purely mechanical case earlier, we can define the total energy function $E:TQ\times\mathbb{R}\rightarrow\mathbb{R}$ for an arbitrary Lagrangian $L$ as follows
\begin{equation}\label{def_E_tot_thermo}
E(\mathbf{q},\dot{\mathbf{q}},S)= \left<\frac{\partial L}{\partial \dot{\mathbf{q}}},  \dot{\mathbf{q}}\right>- L(\mathbf{q}, \dot{\mathbf{q}},S).
\end{equation}
Then, along the solution curve of \eqref{simple_systems}, we have
\[
\frac{d}{dt} E = \left\langle \frac{d}{dt}\frac{\partial L}{\partial \dot{\mathbf{q}}}- \frac{\partial L}{\partial \mathbf{q}}, \dot{ \mathbf{q} } \right\rangle  - \frac{\partial L}{\partial S}\dot S= \left\langle \mathbf{F} ^{\rm ext}( \mathbf{q}, \dot{ \mathbf{q}}),  \dot{ \mathbf{q} } \right\rangle = P^{\rm ext}_W,
\]
in which $P_W^{\rm ext}$ denotes the mechanical power due to the external forces $\mathbf{F} ^{\rm ext}$ that are imposed on the system. This is the first law \eqref{1st_law} for the thermodynamic system \eqref{simple_systems}.

\paragraph{Entropy balance.}
From the last equation in \eqref{simple_systems} the rate of entropy production of the system is
\[
\dot S= -\frac{1}{T}\left< \mathbf{F} ^{\rm fr}(\mathbf{q},\dot{\mathbf{q}},S),\dot{\mathbf{q}}\right>.
\]
The second law means that $ \dot  S$ is always positive, and it follows that the friction force must be dissipative. It also follows that the phenomenological relation is given by $F ^{\rm fr}_{i}=-\lambda_{ij} \dot{q}^{j}$, where the state functions $\lambda_{ij}= \textcolor{black}{ \lambda _{ij}( \mathbf{q} , \dot{\mathbf{q}}, S)},\; i,j=1,...,n$ are usually determined by experiments, with the symmetric part of the matrix $\lambda_{ij}$ positive semi-definite. 
\medskip

We show our variational formulation with the case of a thermo-mechanical and a thermo-electrical system, in which the unifying character of the formulation is illustrated.

\begin{figure}[h]
\vspace{-0.3cm}
\begin{center}
%\hspace{2cm}
\includegraphics[scale=0.55]{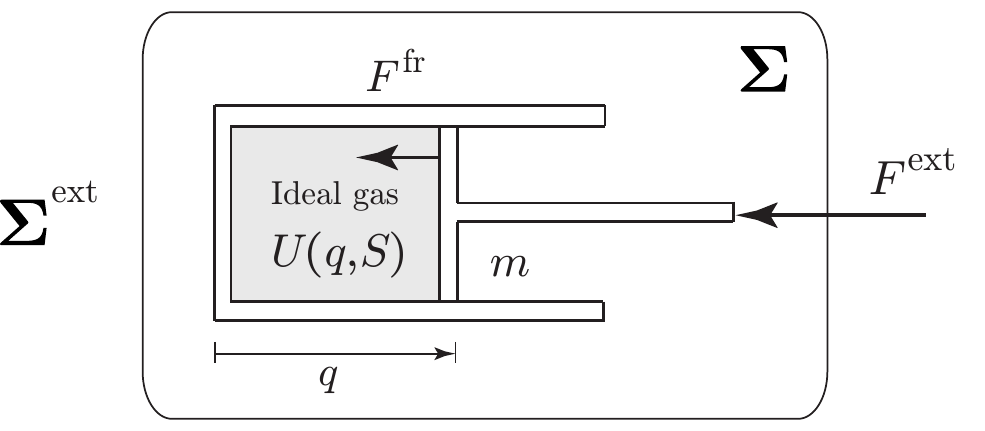}
\caption{System of an ideal gas confined in a cylinder by a piston}
\label{piston}
\end{center}
\vspace{-1.2cm}
\end{figure}

\paragraph{Example: an ideal gas in a cylinder with movable piston.} Consider a gas confined by a piston in a cylinder as in Figure \ref{piston}. Let $m$ be the mass of the piston and $q \in Q= \mathbb{R} $ be the displacement of the piston.
Let $\mathsf{U}(S,V,N)$ be the internal energy of the ideal gas, $N=N_0$ the number of moles which is assumed to be constant, $V= Aq$  the volume, and $A$ the constant area of the cylinder. Then the Lagrangian $L:TQ \times \mathbb{R} \rightarrow \mathbb{R} $ is given by $L(q, \dot  q, S)= \frac{1}{2} m \dot  q ^2 - U(q,S)$, where $U(q,S):= \mathsf{U}(S,V= Aq, N_0)$. Note that we have $ \frac{\partial U}{\partial S} (q,S)= T(q,S)$ the temperature and $ \frac{\partial U}{\partial q}(q,S)=- p(q,S)A$ with $p$ being the pressure. Note also that the friction force is given by using the phenomenological coefficient $\lambda (q, S) \geq 0$ as $F^{\rm fr}(q, \dot  q, S)=- \lambda (q, S)  \dot  q$. We assume that the system is subject to an external force $F^{\rm ext}$. From the variational formulation \eqref{LdA_thermo_simple}--\eqref{CV_simple} applied to this Lagrangian (with $ \Delta _Q=TQ$, i.e., there are no constraints), the equations of motion given by \eqref{simple_systems} yield
\[
m \ddot q = p(q,S) A + F^{\rm ext}- \lambda (q,S) \dot  q, \qquad T(q, S) \dot  S = \lambda (q,S) ^2 \dot  q ^2.
\]
These equations of motion are consistent with those developed in \cite{Gr1999}. \textcolor{black}{Note that more general phenomenological expressions, such as $F^{\rm fr}(q, \dot  q, S)=- \lambda (q, S)f(| \dot  q|)  \dot  q$, where $\lambda (q, S) \geq 0$ and $f(x)\geq 0$ is some increasing scalar function of its argument, are also possible and consistent with the second law.}
\paragraph{Example: circuits with entropy production.}
Let us consider the L-C-R circuit with voltage source $V$ in Figure \ref{fig_INT_LCR} and recall that the configuration space for the charges is $Q =\mathbb{R}^4$. The Lagrangian $L: TQ\times \mathbb{R} \to \mathbb{R}$ of the system is given by
$
L(\mathbf{q},\mathbf{\dot{q}},S)=\frac{1}{2}L (\dot{q}_L)^2-\frac{1}{2C}(q_C)^2-U(\mathbf{q},S),
$
where $U(\mathbf{q},S)$ is the internal energy. Recall that $\Delta_Q \subset TQ$ is a constraint distribution associated with the KCL constraints in \eqref{KCL_const} and also that $F^{\rm fr}(\mathbf{q}, \mathbf{\dot{q}}, S):=V=-R(\mathbf{q},S)f_R$, where $R(\mathbf{q},S)=R$ is a resistive constant. The equations of motion are given by \eqref{simple_systems}, which take the form
\[
\begin{split}
L\ddot{q}_L=V+Rf_R,\quad  R\dot{q}_R=\frac{1}{C}q_C,\quad \dot{q}_C=\dot{q}_L-\dot{q}_R,\quad  \dot{S}=\frac{1}{T}R\dot{q}_R^2,
\end{split}
\]
where $T=\frac{\partial U}{\partial S}$ denotes the temperature and $e_V=-V$.

\subsection{Variational formulation for non-simple thermodynamic systems}\label{subsec_2c}

%\todo{FGB: I rewrote a little bit here by changing the indices $A \rightarrow k$, since we have the indices $k$ elsewhere. Also a slightly extend the variational formulation so that it exactly corresponds to what we get later by interconnection (i.e. I stay with the $q_k$ and $ \Sigma _Q$, whereas in our Part 1 paper we already interconnected before taking the variations).
%
%At the end I added informations about how to connect with Prigogine equations.}

Consider a general finite dimensional non-simple thermodynamic system $\boldsymbol{\Sigma}$ where we assume that $\boldsymbol{\Sigma}$ can be decomposed into $P$ subsystems $\boldsymbol{\Sigma}_k, \; k=1,...,P$, i.e., $\boldsymbol{\Sigma}=\cup_{k=1}^P\boldsymbol{\Sigma}_k$. For brevity, suppose that each subsystem $\boldsymbol{\Sigma}_k$ has a single compartment with a single entropy $S_{k} \in \mathbb{R}$ and with mechanical variables $\mathbf{q}_{k} \in Q_{k}$. Hence the state variables for each $\boldsymbol{\Sigma}_k$ are
\begin{equation}\label{list_variables}
(\mathbf{q}_{k},\dot{\mathbf{q}}_{k}, S_k,) \in TQ_{k} \times \mathbb{R}, \quad k=1,...,P.
\end{equation}
We write $L_k:TQ_{k}\times \mathbb{R} \rightarrow  \mathbb{R}$ the Lagrangian of the $k$-th subsystem $\boldsymbol{\Sigma}_k$, $k=1,...,P$. We assume that there are nonholonomic constraints given by distributions $ \Delta _{Q_k} \subset TQ_k$ and we also assume that there exists an {\it interconnection constraint} $\Sigma_Q \subset TQ$ among the subsystems so that we can define the kinematic constraint $\Delta_{Q}=(\Delta_{Q_{1}} \times \cdots \times \Delta_{Q_{P}})\cap \Sigma_Q \subset TQ$.

\begin{figure}[h]
\vspace{-0.3cm}
\begin{center}
%\hspace{2cm}
\includegraphics[scale=0.55]{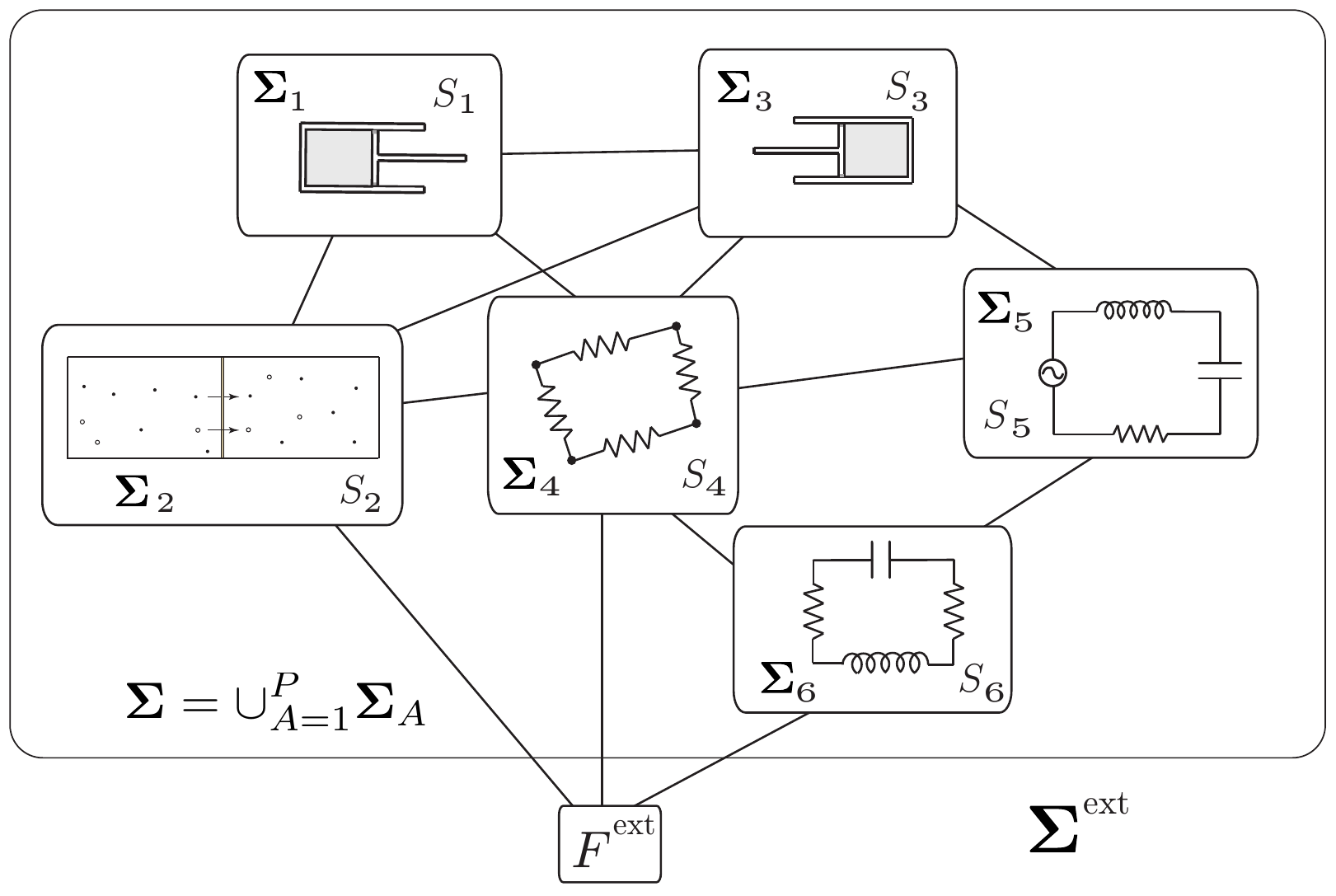}
\caption{Non-simple interconnected system}
\label{Non-simple interconnected system}
\end{center}
\vspace{-1cm}
\end{figure}

Note that the non-simple system $\boldsymbol{\Sigma}$ can be considered as an interconnected system of $P$ simple systems with the state space $T(Q_1 \times ... \times Q_P) \times \mathbb{R}^{P}$, as illustrated in Figure \ref{Non-simple interconnected system}, where each subsystem $\boldsymbol{\Sigma}_k$ is a simple thermodynamic system with the configuration space $Q_k$, which is nothing but a {\it primitive system} in the sense of Kron. This type of non-simple interconnected systems is a natural extension of the class of an {\it interconnected mechanical system} in \cite{JaYo2014} in the sense that it also includes the irreversible processes due to friction and heat conduction.
\medskip

\subsubsection{Variational formulation for systems with friction and heat conduction}\label{subsub_2ci}

Besides the friction forces $ \mathbf{F} ^{\rm fr}_k:TQ_k \times \mathbb{R}  \rightarrow T^*Q_k$ associated to each subsystems, we also need to introduce the entropy fluxes $J_{k\ell}$ associated to the heat exchange between subsystems $ \boldsymbol{\Sigma}_k$ and $ \boldsymbol{\Sigma} _\ell$, where we assume $J_{k\ell}=J_{\ell k}$.

\paragraph{Thermodynamic displacements associated to heat exchange.} An essential ingredient in the variational formulation of thermodynamics is the notion of \textit{thermodynamic displacement associated to an irreversible process}. In general, the thermodynamic displacement associated to an irreversible process $ \alpha $ with thermodynamic force $X^ \alpha $ is a variable $ \Lambda ^ \alpha $ such that its time rate of change satisfies $\dot{ \Lambda }^ \alpha = X^ \alpha $, as defined in \cite{GBYo2017a,GBYo2017b,GBYo2019b}.

For the irreversible process of heat exchange such thermodynamic displacements are given by the variables $\Gamma^k$ which satisfy $\dot \Gamma ^k= T^k$ with $T^k$ the temperature of $ \boldsymbol{ \Sigma }_k$. Note that $\Gamma^k$ is the \textit{thermal displacement} that was used in \cite{GrNa1991}, which was initially coined by \cite{He1884}.  See \S\ref{extension1} for other examples of thermodynamic displacement. The introduction of $\Gamma^k$ is accompanied with the introduction of an entropy variable $\Sigma_k$, usually distinct from $S_k$. The physical meaning of $ \Sigma _k$ will be explained below.

\paragraph{Variational formulation.} Let us consider the variational formulation for a thermodynamic system with friction and heat conduction, in which the total Lagrangian is given by
\[
L: T Q \times \mathbb{R} ^P  \rightarrow \mathbb{R} , \qquad L( \mathbf{q} , \dot { \mathbf{q} },S_1,...,S_P) = \sum_{k=1}^PL_k( \mathbf{q} _k, \dot{ \mathbf{q} }_k, S_k),
\]
with $Q=Q_1 \times ... \times Q_P$.

\begin{theorem}[Variational formulation for non-simple thermodynamic systems]\label{thm_nonsimple_thermodynamic}
The following statements are equivalent:
\begin{itemize}
\item[(i)] The curves $\mathbf{q}(t)=( \mathbf{q} _1(t),..., \mathbf{q} _P(t))$ and $S_1(t),..., S_P(t)$  are critical for the \textit{action functional}
\begin{equation}\label{LdA_thermo_nonsimple} 
\delta \int_{t _1 }^{ t _2} \left[ L(\mathbf{q} , \dot{\mathbf{q}} , S_1,...,S_P)+ \sum_{k=1}^P\dot{\Gamma }^k( S_k- \Sigma  _k)\right] {\rm d}t  =0,
\end{equation}
subject to the \textit{kinematic constraints}
\begin{equation}\label{CK_nonsimple} 
\dot{\mathbf{q}} \in \Delta_{Q}(\mathbf{q}), \qquad \frac{\partial L}{\partial S_k}\dot \Sigma _k  =  \left< \mathbf{F}_k ^{\rm fr}, \dot{\mathbf{q}}_k \right>+\sum_\ell J_{k\ell} \dot  \Gamma ^\ell,\;k=1,...,P
\end{equation}
and for variations subject to the  \textit{variational constraints}
\begin{equation}\label{CV_nonsimple} 
\delta{\mathbf{q}} \in \Delta_{Q}(\mathbf{q}), \qquad \frac{\partial L}{\partial S_k}\delta \Sigma _k=  \left< \mathbf{F} ^{\rm fr}_k, \delta \mathbf{q}_k \right>+\sum_\ell J_{k\ell} \delta   \Gamma ^\ell,\;k=1,...,P
\end{equation}
with $ \delta \mathbf{q}(t_1)=\delta \mathbf{q}(t_2)=0$.
\medskip 

\item[(ii)] The curves $\mathbf{q}(t)=( \mathbf{q} _1(t),..., \mathbf{q} _P(t))$ and $S_1(t),..., S_P(t)$ are solutions of the system of equations
\begin{equation}\label{nonsimple_systems} 
\left\{
\begin{array}{l}
\displaystyle\vspace{0.2cm}\frac{d}{dt}\frac{\partial L}{\partial \dot{\mathbf{q}}}- \frac{\partial L}{\partial \mathbf{q}} - \mathbf{F} ^{\rm fr}(\mathbf{q}, \dot{\mathbf{q}}, S_1,...,S_P) \in \Delta_Q^{\circ}(\mathbf{q}),\\
\displaystyle \dot{\mathbf{q}} \in \Delta_Q(\mathbf{q}), \qquad \frac{\partial L}{\partial S_k}\dot S_k=  \left< \mathbf{F}_k ^{\rm fr}, \dot{\mathbf{q}}_k \right> + \sum_\ell J_{k\ell} \left( \frac{\partial L}{\partial S_k}- \frac{\partial L}{\partial S_\ell}\right),\;k=1,...,P.
\end{array} \right.
\end{equation} 
\end{itemize}
We recall that the kinematic constraint $ \Delta _Q$ on the total configuration space $Q= Q_1 \times .... \times Q_P$ is built from the nonholonomic constraints $ \Delta _{Q_k}$ within each subsystem $\boldsymbol{\Sigma} _k$ and from the interconnection constraints $ \Sigma _Q$ among each subsystems as $\Delta_{Q}=(\Delta_{Q_{1}} \times \cdots \times \Delta_{Q_{P}})\cap \Sigma_Q \subset TQ$.
\end{theorem}

The derivation of the final system of equations \eqref{nonsimple_systems} from \eqref{LdA_thermo_nonsimple}--\eqref{CV_nonsimple}  is obtained by noting the two following conditions associated to the variations $ \delta S_k$ and $\delta \Gamma ^k$ when applying \eqref{LdA_thermo_nonsimple}--\eqref{CV_nonsimple}:
\begin{equation}\label{Gamma_Sigma} 
\delta S_k: \;\; \frac{\partial L}{\partial S_k} =- \dot  \Gamma ^k, \qquad \delta \Gamma ^k:\;\; \dot S_k= \dot \Sigma _k+ \sum_\ell J_{k\ell}.
\end{equation} 
The first condition states that $ \Gamma ^k$ indeed corresponds to the thermal displacement, while the second condition defines the entropy variable $ \Sigma _k$ in terms of $S_k$ and the entropy fluxes. The second condition is related to the Prigogine equation 
\begin{equation}\label{Prigogine}
{\rm d} S= {\rm d} _iS + {\rm d} _eS
\end{equation}
as we shall see later. We note that $\dot \Gamma ^k$ and $\dot \Sigma _k$ ultimately cancel in the final form of the equations.

In the first equation of \eqref{nonsimple_systems} we have defined $ \mathbf{F} ^{\rm fr}=(   \mathbf{F} ^{\rm fr}_1,...,  \mathbf{F} ^{\rm fr}_P)$. We recall that the distribution $ \Delta _Q \subset TQ$ is defined by $ \Delta _Q= ( \Delta _{Q_1} \times ... \times \Delta _{Q_P}) \cap \Sigma _Q$, hence the first and second equations can be equivalently written as
\begin{equation}\label{nonsimple_systems_rewriting}
\begin{aligned} 
&(\dot{\mathbf{q} }_1,..., \dot{\mathbf{q}}_P) \in \Sigma _Q, \quad \qquad \dot{\mathbf{q}} _k \in \Delta _{Q_k}( \mathbf{q} _k), \quad \qquad \frac{d}{dt}\frac{\partial L}{\partial \dot{\mathbf{q}}_k}- \frac{\partial L}{\partial \mathbf{q}_k} - \mathbf{F} _k^{\rm fr} - \mathbf{F} _k \in \Delta _{Q_k}^\circ,
\end{aligned}
\end{equation}
$k=1,...,P$, for some interaction forces $( \mathbf{F} _1,..., \mathbf{F} _k)  \in \Sigma _Q^\circ$.

\paragraph{Energy balances and the first law.} Let us consider the energy $E_k:TQ_k \times \mathbb{R} \rightarrow \mathbb{R} $ of the $k$-th subsystem $ \boldsymbol{\Sigma} _k$ as
\begin{equation}\label{def_Ek}
E_k(\mathbf{q}_k,\mathbf{v}_k,S_k)= \left<\frac{\partial L}{\partial \mathbf{v}_k},  \mathbf{v}_k\right>- L(\mathbf{q}_k, \mathbf{v}_k,S_k).
\end{equation}
From the evolution equations \eqref{nonsimple_systems} one gets the energy balance
\[
\frac{d}{dt} E_k = \left\langle \mathbf{F} _k, \dot{ \mathbf{q} }_k \right\rangle + \sum_\ell J_{k\ell} \left( \frac{\partial L}{\partial S_\ell}- \frac{\partial L}{\partial S_k}\right)= P^{{\rm int} \rightarrow k}_W + P_H^{\ell \rightarrow k},
\] 
which allows to identify the mechanical power $P^{{\rm int} \rightarrow k}_W$ associated to the work done on the subsystem $ \boldsymbol{\Sigma}_k$ arising from the interconnection within the total system $ \boldsymbol{ \Sigma }$, as well as the power $P_H^{\ell \rightarrow k}$ associated to the heat transfer  from $\boldsymbol{ \Sigma }_\ell$ to $\boldsymbol{ \Sigma }_k$.
This is nothing else than the first law associated to the $k$-th subsystem, where we note that the work of the friction force $ \mathbf{F} ^{\rm fr}_k$ does not appear since it is an internal force associated to an irreversible process within the $k$-th subsystem.

The total energy balance for $E= \sum_k E_k$ is found as
\[
\frac{d}{dt} E= \sum_k\left\langle \mathbf{F} _k, \dot{ \mathbf{q} }_k \right\rangle + \sum_{k,\ell}J_{k\ell} \left( \frac{\partial L}{\partial S_\ell}- \frac{\partial L}{\partial S_k}\right)=0,
\]
where we note that the first term vanishes due to the interconnection condition $(\dot{\mathbf{q} }_1,..., \dot{\mathbf{q}}_P) \in \Sigma _Q$ and $( \mathbf{F} _1,..., \mathbf{F} _k)  \in \Sigma _Q^\circ$.
In fact the total energy is preserved since the system $ \boldsymbol{\Sigma} $ is isolated.

\paragraph{Entropy balances and the second law.} Recalling that $ -\frac{\partial L}{\partial S_k} =T^k$ is the temperature of the $k$-th subsystem, its entropy balance is found from \eqref{nonsimple_systems} as
\[
\dot  S_k= - \frac{1}{T^k} \left< \mathbf{F}_k ^{\rm fr}, \dot{\mathbf{q}}_k \right> + \sum_\ell J_{k\ell} \frac{T^k-T^\ell}{T^k}.
\]
It is important to note that the second law does not implies $ \dot  S_k\geq 0$ since the $k$-th subsystem is not adiabatically closed. To apply the second law one has to identify the rate of internal (as opposed to total) entropy production.
In our variational approach, the rate of internal entropy production is identified with $ \dot  \Sigma _k$ and from the second condition \eqref{Gamma_Sigma}, one directly obtains the rate of internal entropy production as
\[
\dot  \Sigma _k = - \frac{1}{T^k} \left< \mathbf{F}_k ^{\rm fr}, \dot{\mathbf{q}}_k \right> - \sum_\ell J_{k\ell} \frac{T^\ell}{T^k} .
\] 
The second law implies $- \frac{1}{T^k} \left< \mathbf{F}_k ^{\rm fr}, \dot{\mathbf{q}}_k \right> \geq 0$ and $\sum_\ell J_{k\ell} \frac{T^\ell}{T^k}\geq 0$, thus forcing $ \mathbf{F} _k^{\rm fr}$ to be dissipative forces and $J_{k\ell}$ to be a non-positive state function.

Because the entropy is considered to be an extensive variable, the total entropy of the system is given by $S=\sum_{k=1}^PS_k$. By summing the entropy balances of each subsystem we get
\begin{equation}\label{internal_entropy_production}
\dot{S}=-\sum_{k=1}^P\frac{1}{T^k} \left< \mathbf{F} _k^{\rm fr}, \dot{\mathbf{q} }_k \right> +
\sum_{k<\ell}^{K}J_{k\ell }\left(\frac{1}{T^k}-\frac{1}{T^\ell} \right) (T^k-T^\ell).
\end{equation}
Since the total system $ \boldsymbol{\Sigma }$ is isolated, the second law imposes $ \dot  S \geq 0$. This is indeed the case from the conditions on $ \mathbf{F} ^{\rm fr}_k$ and $J_{k\ell}$ already found earlier.
\medskip

In terms of the Prigogine equation \eqref{Prigogine} these results can be summarized as follows:
\begin{align*} 
\boldsymbol{\Sigma} _k&:\; {\rm d} S_k = {\rm d} _i S_k + {\rm d} _eS_k, \quad {\rm d} _i S_k \geq 0,\;\; {\rm d} _eS_k \text{ no sign},\\
\boldsymbol{\Sigma} &:\; {\rm d} S= {\rm d} _i S + {\rm d} _eS, \quad \quad \;\;\,{\rm d} _i S \geq 0, \;\; {\rm d} _eS=0,
\end{align*} 
where our variational formulation gives the concrete expressions for each of these quantities as
\begin{align*} 
\frac{{\rm d} _i S_k}{ {\rm d} t} &= \dot \Sigma _k = - \frac{1}{T^k} \left< \mathbf{F}_k ^{\rm fr}, \dot{\mathbf{q}}_k \right> - \sum_\ell J_{k\ell} \frac{T^\ell}{T^k}, \qquad \frac{{\rm d} _e S_k}{ {\rm d} t} = \sum_\ell J_{k\ell},\\
\frac{ {\rm d}S}{ {\rm d} t} &= \dot  S = -\sum_{k=1}^P\frac{1}{T^k} \left< \mathbf{F} _k^{\rm fr}, \dot{\mathbf{q} }_k \right> +
\sum_{k<\ell}^{K}J_{k\ell }\left(\frac{1}{T^k}-\frac{1}{T^\ell} \right) (T^k-T^\ell).
\end{align*} 
\textcolor{black}{In particular, the equations  ${\rm d} _i S_k= \dot \Sigma _k  {\rm d} t$ and $ {\rm d} S_k= \dot  S_k {\rm d} t$ explicitly show the different physical meanings of the two entropy variables $ \Sigma _k$ and $S_k$ used in the variational formulation.}

\subsubsection{Extensions to matter exchange, chemical reactions, and open systems}\label{extension1}

The variational formulation for non-simple systems developed in \S\ref{subsub_2ci} can be extended to the case in which each subsystem contains several chemical species undergoing chemical reactions, with possible diffusion of the species between the subsystems in addition to heat exchange. This is achieved by introducing the thermodynamic displacements associated to matter transport and chemical reactions. From its general definition as a time integral of the thermodynamic force, the thermodynamic displacement associated to the matter transport of a chemical species $k$ is given by the variable $W^k$ such that its time rate of change is the chemical potential $ \mu ^k$ of this species, namely $ \dot  W^k= \mu ^k$. Similarly, the thermodynamic displacement associated to a chemical reaction is such that its time rate of change is the affinity of the reaction. With these concepts, the variational formulation \eqref{LdA_thermo_nonsimple}--\eqref{CV_nonsimple} extends to these cases, while keeping the same structure, see \cite{GBYo2017a}, \cite{GBYo2019b}.
An appropriate extension of the variational formulation \eqref{LdA_thermo_nonsimple}--\eqref{CV_nonsimple} also allows the treatment of open systems that exchange heat, matter, and kinetic energy with their surroundings, in which case the constraint becomes affine and explicitly time dependent, \cite{GBYo2018b}.

\subsubsection{Extensions to continuum systems}\label{extension2}
It is also possible to extend \eqref{LdA_thermo_nonsimple}--\eqref{CV_nonsimple} to continuum systems in order to treat for instance the case of multicomponent reacting heat conducting viscous fluids, \cite{GBYo2017b,GBYo2019b}. The structure of the variational formulation remains the same, with the phenomenological and variational constraints related as above, and recovers the Hamilton principle of continuum mechanics in absence of irreversibility.
This approach is especially useful as a modelling tool for the derivation of thermodynamically consistent models, especially in systems involving constraints in their variational formulations, such as semi-incompressible fluids, \cite{ElGB2021} or porous media \cite{GBPu2022}, as well as for the derivation of thermodynamically consistent numerical discretization \cite{GBYo2018a,GaGB2022}.

\color{black}

\section{Interconnection of thermodynamic systems}\label{sec_4}

Here we illustrate the variational formulation of {\it interconnected thermodynamic systems} by extending the idea of interconnected systems in mechanics. This approach is crucial when studying {\it multiphysical systems}, their interconnection, and their thermodynamic consistency. We start by reviewing the case of mechanics, and then develop the case of simple as well as non-simple thermodynamic systems.

\color{black} 
\subsection{Variational formulation for interconnected mechanical systems}\label{SubSubSec_VF_IntConnection}

The variational formulation for an interconnected system in mechanics was developed by \cite{JaYo2014}. Let us see how it can be formulated for the case in which a mechanical system with configuration manifold $Q=Q_{1} \times Q_{2}\ni  \mathbf{q} =( \mathbf{q} _1, \mathbf{q} _2)$ is decomposed into two mechanical systems $k=1,2$. Here we suppose also that each $k$-th mechanical system, called a primitive system, has constraints $\Delta _{Q_k} \subset TQ_k$ in which $Q_k$ denotes a configuration manifold of the $k$-th primitive system.

The variational formulation for each subsystem $k=1,2$ is given by
\begin{equation}\label{VF_primitive} 
\delta  \int_{t_0}^{t_1}  L_k( \mathbf{q}_k , \dot {\mathbf{q}}_k )  {\rm d} t + \int_{t_0}^{t_1} \left\langle \mathbf{F}_k ( \mathbf{q} , \dot{\mathbf{q}}),  \delta \mathbf{q}_k \right\rangle  {\rm d} t=0, \quad \delta \mathbf{q} _k \in \Delta _k( \mathbf{q}_k ), \quad\text{for}\quad  k=1,2
\end{equation} 
with the condition $\dot{\mathbf{q} }_k \in \Delta _{Q_k}( \mathbf{q} _k) \subset T_{\mathbf{q} _k}Q_{k}$, where $\mathbf{F}_k: TQ \to T^{\ast}Q_k$ is the interaction force at the $k$-th boundary. We note that, unlike $ L_k$,  the interaction force $\mathbf{F}_k$ is defined on $TQ$ and not on $TQ_k$ only. One gets from \eqref{VF_primitive} the equations of motion for each primitive system as
\begin{equation}\label{EL_kth} 
\frac{d}{dt} \frac{\partial L_k}{\partial \dot{\mathbf{q}}_k}- \frac{\partial L_k}{\partial \mathbf{q}_k} - \mathbf{F} _k\in \Delta_{Q_k}^\circ(\mathbf{q}_k), \qquad \dot{\mathbf{q}}_k \in \Delta _{Q_k}( \mathbf{q}_k).
\end{equation}

The interconnection of the two mechanical systems is given by imposing some distribution $ \Sigma _Q \subset T(Q_1 \times Q_2)$ such that
\begin{equation}\label{interconnection_cond} 
( \dot {\mathbf{q} }_1, \dot {\mathbf{q}} _2) \in \Sigma _Q, \qquad ( \mathbf{F} _1, \mathbf{F}_2) \in \Sigma _{Q}^ \circ .
\end{equation}

The dynamics of the interconnected system, given by \eqref{EL_kth} together with the interconnection condition \eqref{interconnection_cond}, is equivalently provided by the interconnected variational formulation
\begin{equation}\label{VF_interconnected}
\delta  \int_{t_0}^{t_1} \left[ L_1( \mathbf{q}_1 , \dot {\mathbf{q}}_1 ) + L_2( \mathbf{q}_2 , \dot {\mathbf{q}}_2)  \right]  {\rm d} t=0, \quad (\delta \mathbf{q} _1, \delta \mathbf{q} _2) \in \Delta _Q( \mathbf{q}_1, \mathbf{q} _2 ),
\end{equation}
with the condition $(\dot{\mathbf{q} }_1, \dot{\mathbf{q}} _2) \in \Delta _Q( \mathbf{q}_1, \mathbf{q} _2 )$, where we have defined the new distribution
\begin{equation}\label{def_Delta_Q} 
\Delta _Q:= \left(  \Delta _{Q_1} \times \Delta _{Q_2} \right)  \cap \Sigma _Q.
\end{equation} 
Indeed, \eqref{VF_interconnected} gives
\[
\left( \frac{d}{dt} \frac{\partial L_1}{\partial \dot{\mathbf{q}}_1}- \frac{\partial L_1}{\partial \mathbf{q}_1} , \frac{d}{dt} \frac{\partial L_2}{\partial \dot{\mathbf{q}}_2}- \frac{\partial L_2}{\partial \mathbf{q}_2} \right) \in \Delta  _Q^\circ, \qquad ( \dot {\mathbf{q} }_1, \dot {\mathbf{q}} _2) \in \Delta  _Q( \mathbf{q} _1, \mathbf{q} _2)
\]
which is clearly equivalent to \eqref{EL_kth} and \eqref{interconnection_cond} by using \eqref{def_Delta_Q}. 

\begin{remark}[On primitive systems and interaction forces]\rm
In the above, the primitive system seems to be similar to a system with external force, but strictly speaking it is not the same for two reasons. The first reason is that the $k$-th interaction force $ \mathbf{F} _k^{\rm int}: TQ \to T^{\ast}Q_k$ is defined as a map from the tangent bundle $TQ$ of the {\it total space} $Q$ to the cotangent bundle of the $k$-th manifold. This means that the primitive system itself is just a piece that is torn apart from the original system and it makes no physical sense by itself alone. In other words, it makes sense as a disconnected piece of the original system that has to be interconnected with other primitive systems to reconstruct the original system. Then, the interconnected system built from the primitive systems does work as a physical system if we correctly interconnect them. This is a natural way for humans to model complicated systems. To reconstruct the full system, we need to assemble them and each part has an appropriate boundary with some other parts that must be bonded. This ingredient to bond with other parts must be the interaction forces and velocities in our case (the modeler knows how to interconnect them because he knows how they were cut from the original system). If the primitive system is a system with an external force, as we have above, then it must make sense by itself alone and then the force must be defined on $TQ_k$, i.e., $\mathbf{F} _k^{\rm ext}: TQ_k \to T^{\ast}Q_k$. The second reason is that the interaction forces $\mathbf{F} _k^{\rm int}$ must satisfy the interconnection constraints, see \eqref{interconnection_cond}, and hence they are rather {\it constraint forces} than external forces.
\end{remark}

\begin{figure}[h] %  figure placement: here, top, bottom, or page
\vspace{-0.3cm}
\centering
\includegraphics[scale=.8]{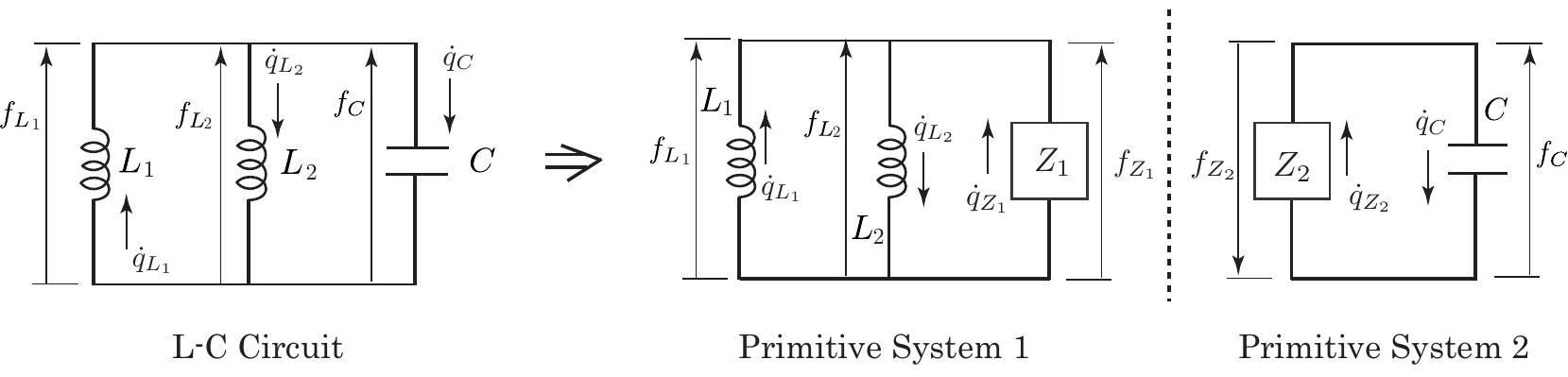}
\caption{L-C Circuit}
\label{vrlc_circuit_pic}
\vspace{-0.5cm}
\end{figure}

\paragraph{Example: L-C circuit.} Let us illustrate the variational setting for interconnected systems given above with an example of the L-C circuit that  is decomposed into two disconnected primitive systems as in  Figure \ref{vrlc_circuit_pic}.  Let $Z_{1}$ and $Z_{2}$ denote the external ports resulting by tearing the original system. To establish the original circuit in Figure \ref{vrlc_circuit_pic}, the external ports are interconnected by equating currents across them.

For this example, the primitive system 1 has the configuration space $Q_{1}=\mathbb{R}^3$ with local coordinates
$\mathbf{q}_1 =(q_{L_1}, q_{L_2}, q_{Z_1})$, where $q_{L_1}, q_{L_2}$ and $q_{Z_{1}}$ denote respectively the charges associated to the inductor $L_1$, inductor $L_2$ and port $Z_{1}$. The primitive system 2 has the configuration space $Q_{2}= \mathbb{R}^2$ with local coordinates $\mathbf{q}_{2}=(q_{Z_2},q_C)$, where $q_{Z_2}$ is the charge through the port $Z_2$ and $q_C$ is the charge stored in the capacitor. We have $Q= Q_1 \times Q_2 \ni \mathbf{q} =( \mathbf{q} _1, \mathbf{q} _2)=(q_{L_1}, q_{L_2}, q_{Z_1}, q_{Z_2},q_C)$.

\paragraph{Primitive system 1.}
Kirchhoff's circuit law (KCL) is enforced  by applying the constraint distribution $\Delta_{Q_{1}} \subset TQ_{1}$ given by
\[
\Delta_{Q_{1}}(\mathbf{q}_{1}) = \{ \dot{\mathbf{q}} _{1}=( \dot  q_{L_1},  \dot  q_{L_2}, \dot  q_{Z_1}) \in T_{\mathbf{q}_{1}}Q_1 \mid  \dot  q_{L_1} - \dot  q_{L_2} - \dot  q_{Z_1} = 0 \}
\]
for each $\mathbf{q}_1 =(q_{L_1}, q_{L_2}, q_{Z_1}) \in Q_{1}$, where $\dot{\mathbf{q}}_{1}=(\dot  q_{L_1}, \dot  q_{L_2} \dot  q_{Z1})$ denotes the current vector at each $\mathbf{q}_{1}$. The corresponding Kirchhoff's voltage law (KVL) constraint is described by its annihilator $\Delta_{Q_{1}}^\circ$ which reads
\[
\Delta_{Q_{1}}^{\circ}(\mathbf{q}_{1}) = \{  \mathbf{f} _{1}=(f_{L_1}, f_{L_2}, f_{Z_1}) \in T_{\mathbf{q}_{1}}^{\ast}Q_1 \mid  f_{L_1} =- f_{L_2}, -f_{L_1}= f_{Z_1}  \}
\]
for each $\mathbf{q}_1 =(q_{L_1}, q_{L_2}, q_{Z_1}) \in Q_{1}$. 
The Lagrangian for the primitive circuit 1 is $\mathcal{L}_{1}:TQ_{1} \rightarrow \mathbb{R} $, 
$
\mathcal{L}_1(\mathbf{q}_1, \dot{\mathbf{q}}_1) = \frac{1}{2} L_{1} \dot  q_{L_{1}}^2+  \frac{1}{2} L_{2}\dot  q_{L_{2}}^2.
$
The voltage associated to the port $Z_1$ is denoted by $f_{Z_{1}}(\mathbf{q}, \dot{\mathbf{q}})dq_{Z_{1}}$, giving the interaction voltage field $ \mathbf{F}_{1}: TQ \to T^{\ast}Q_{1}$ as
$
\mathbf{F}_{1}(\mathbf{q},\dot{\mathbf{q}} )=(q_{L_{1}}, q_{L_{2}}, q_{Z_1},0,0,f_{Z_{1}}(\mathbf{q}, \dot{\mathbf{q}})),
$
with $(\mathbf{q}, \dot{\mathbf{q}}) = (q_{L_{1}}, q_{L_{2}}, q_{Z_1}, q_{Z_2}, q_{C}, \dot  q_{L_{1}}, \dot  q_{L_{2}}, \dot  q _{Z_1}, \dot  q_{Z_2}, \dot  q_{C}) \in TQ$.

With the Lagrangian $ \mathcal{L} _1$, constraint $ \Delta _{Q_1}$ and interaction force $ \mathbf{F}_1$, the equations of motion \eqref{EL_kth} for the primitive circuit 1 are
\begin{equation}\label{primCir1}
L_{1}\ddot{q}_{L_{1}}=-\lambda_{1}, \qquad L_{2}\ddot{q}_{L_{2}}=-\lambda_{1},\qquad
f_{Z_{1}}=\lambda_{1}, \qquad \dot  q_{L_1} - \dot  q_{L_2} - \dot  q_{Z_1} = 0.
\end{equation}
These equations of motion are well defined for each given curve $(\mathbf{q}_2(t), \dot{ \mathbf{q} }_2(t) ) \in TQ_2$.

\paragraph{Primitive system 2.}
The KCL space is given by
\[
	\Delta_{Q_{2}}(\mathbf{q}_{2}) = \{ \dot{\mathbf{q}}_2=(\dot  q_{Z_2}, \dot  q_C) \in T_{\mathbf{q}_{2}}Q_2 \mid \dot  q _C - \dot  q_{Z_2} = 0 \}
\]
for each $\mathbf{q}_{2}=(q_{Z_2},q_C) \in Q_{2}$, hence the KVL space described by the annihilator $\Delta_2^\circ(\mathbf{q}_{2})$ is found as
\[
	\Delta^{\circ}_{Q_{2}}(\mathbf{q}_{2}) = \{ \mathbf{f}_2=(f_{Z_2},f_C) \in T_{\mathbf{q}_{2}}^{\ast}Q_2 \mid -f_C = f_{Z_2} \}.
\]
The Lagrangian for the primitive circuit 2 is $\mathcal{L}_2:TQ_{2} \to \mathbb{R}$, 
$
\mathcal{L}_2 = \frac{1}{2C} q_{C}^2.
$
Given the voltage $f_{Z_{2}}(\mathbf{q},\dot{\mathbf{q}})dq_{Z_{2}}$ associated to the port $Z_2$, the interaction voltage field follows as
$
F_{2}(\mathbf{q},\dot{\mathbf{q}})=(q_{Z_2},q_C,f_{Z_{2}}(\mathbf{q},\dot{\mathbf{q}}),0).
$

With the Lagrangian $ \mathcal{L} _2$, constraint $ \Delta _{Q_2}$ and interaction force $ \mathbf{F}_2$, 
the equation of motion \eqref{EL_kth} of the primitive system 2 are
\begin{equation}\label{primCir2}
\begin{split}
\dot  q_C - \dot  q_{Z_2} = 0, \quad -\frac{q_{C}}{C}= f_{Z_2}(\mathbf{q},\dot{\mathbf{q}}).
\end{split}
\end{equation}
These equations of motion are well defined for each given curve $(\mathbf{q}_{1}(t), \dot{ \mathbf{q} }_{1}(t) ) \in TQ_1$.

\paragraph{The interconnected system.} The interconnection of the two primitive systems is given by imposing the equality of currents across the ports, which results in the distribution $ \Sigma _Q$ on $Q = Q_1 \times Q_2$ given by
\[
\Sigma_{Q} = \{ (\dot  q_{L_{1}},\dot  q_{L_{2}},\dot  q_{Z_{1}},\dot  q_{Z_{2}}, \dot  q_C) \in TQ \mid \dot  q_{Z_1}=\dot  q_{Z_2} \},
\]
with the annihilator
$$
\Sigma_{Q}^{\circ} = \{ (0,0,f_{Z_{1}},f_{Z_{2}},0) \in T^{\ast}Q \mid f_{Z_1}+f_{Z_2}=0  \}.
$$
In this way, the current (velocity) $\dot{\mathbf{q}}_{\rm int}=(\mathbf{q}_{L_{1}}, \mathbf{q}_{L_{2}}, \mathbf{q}_{Z_{1}},\mathbf{q}_{Z_{2}}, \mathbf{q}_C)$ and voltage (force) $\mathbf{f}_{\rm int}=(0,0,f_{Z_{1}},f_{Z_{2}},0)$ at the boundaries must satisfy the constraint $(\dot{\mathbf{q}}_{\rm int},\mathbf{f}_{\rm int}) \in \Sigma_{Q} \oplus \Sigma_{Q}^{\circ}$, see \eqref{interconnection_cond}.

The equations of motion for the interconnected dynamical system are provided by the system of equations for each primitive systems \eqref{primCir1} and \eqref{primCir2} together with the interconnection condition
$\dot  q_{Z_{1}}=\dot  q_{Z_{2}}$ and $f_{Z_{1}}+f_{Z_{2}}=0$.  Hence we finally get the evolution equations
$$
L_{1}\ddot{q}_{L_{1}}= -\frac{q_{C}}{C}, \quad L_{2}(\ddot{q}_{L_{1}}-\ddot{q}_{C})=-\frac{q_{C}}{C}.
$$

As stated in \eqref{VF_interconnected}, these equations of evolution are also obtained by the variational formulation for the interconnected system in which we employ the new distribution $ \Delta _Q= ( \Delta _{Q_1} \times \Delta _{Q_2}) \times \Sigma _Q$ for the variational and kinematic constraints, and we also note that the internal variables $ \mathbf{F}_{k},\,k=1,2$ do not appear in the variational condition.

\subsection{Variational formulation for interconnected simple thermodynamic systems}\label{subsec_4b}

The variational formulation of simple thermodynamic systems has been stated in Theorem \ref{thm_simple_thermodynamic}. Here, we consider the case in which the Lagrangian $L:TQ \times \mathbb{R} \rightarrow \mathbb{R} $ is split into a mechanical and a thermal part as $L( \mathbf{q} , \dot{ \mathbf{q} }, S)= L_{\rm mech}( \mathbf{q} , \dot{ \mathbf{q} }) -U(S)$.

%\todo{FGB: Is it possible to have a figure for this kind of interconnection? as done in Figure \ref{ClosedHeatTransSim}.}

\paragraph{Interconnected simple systems.} Let us describe how the thermodynamic system can be understood via interconnection. We reticulate the system into two primitive systems, namely, a primitive system 1 with a purely mechanical part $L_{\rm mech}( \mathbf{q} ,\dot{ \mathbf{q} })$ and a primitive system 2 with a purely thermal part $U(S)$.
By tearing the system into two parts, the {\it intermediate or boundary} variables $\dot{ \mathbf{q} },\dot{\bar{ \mathbf{q} }}$ and $\mathbf{F} ^{\rm int}, \bar{\mathbf{F}}^{\rm int}$ appear.

\paragraph{Primitive system 1:} For the {\it primitive system 1}, we consider the variational condition
\begin{equation*}
\delta \int_{t _1 }^{ t _2} L_{\rm mech}\left( \mathbf{q} , \dot{ \mathbf{q} }\right) {\rm d} t + \int_{t _1 }^{ t _2} \left\langle \mathbf{F} ^{\rm int}\left( \mathbf{q} , \dot { \mathbf{q} }, \bar{ \mathbf{q} }, \dot{\bar{\mathbf{q}}}, S\right) ,  \delta{ \mathbf{q} } \right\rangle {\rm d} t=0,
\end{equation*}
with $\delta  \mathbf{q} (t_1)=\delta \mathbf{q} (t_2)=0$, yielding the equation
\begin{equation}\label{primitive_syst_1_simple} 
\frac{d}{dt}\frac{\partial L_{\rm mech}}{\partial \dot{ \mathbf{q}}}- \frac{\partial L_{\rm mech}}{\partial  \mathbf{q} }=  \mathbf{F} ^{\rm int}.
\end{equation}

\paragraph{Primitive system 2:} For the {\it primitive system 2}, we consider the variational condition
\begin{equation*}
\delta \int_{t _1 }^{ t _2} \left[-U(S)\right] {\rm d} t + \int_{t _1 }^{ t _2} \left\langle  \bar{\mathbf{F} }^{\rm int}(\mathbf{q} , \dot { \mathbf{q} }, \bar{ \mathbf{q} }, \dot{\bar{\mathbf{q}}}, S) , \delta{\bar{ \mathbf{q} }} \right\rangle {\rm d} t=0,
\end{equation*}
subject to the \textit{phenomenological constraint}
\begin{equation*}
-\frac{\partial U}{\partial S}\dot S  =  \left\langle \mathbf{F} ^{\rm fr}(\bar{ \mathbf{q} },\dot{\bar{ \mathbf{q} }}, S) , \dot{\bar{\mathbf{q} }} \right\rangle ,
\end{equation*}
and for variations which are subject to the \textit{variational constraint}
\begin{equation*}
-\frac{\partial U}{\partial S}\delta S  = \left\langle \mathbf{F} ^{\rm fr}(\bar{\mathbf{q} },\dot{\bar{\mathbf{q} }}, S) ,  \delta{\bar{\mathbf{q} }} \right\rangle 
\end{equation*}
with $\delta{\bar{\mathbf{q} }}(t_1)=\delta{\bar{\mathbf{q} }}(t_2)=0$. 
It gives, the system equations for the primitive system 2 as
\begin{equation}\label{Prim2_SimpleSys}
\left\{
\begin{array}{l}
\displaystyle\vspace{0.2cm}  \bar{ \mathbf{F} }^{\rm int}+ \mathbf{F} ^{\rm fr}=0,
\\
\displaystyle -\frac{\partial U}{\partial S}\dot S=  \left\langle \mathbf{F} ^{\rm fr}(\bar{ \mathbf{q} },\dot{\bar{ \mathbf{q} }}, S), \dot{\bar{ \mathbf{q} }} \right\rangle .
\end{array} 
\right.
\end{equation} 

\paragraph{Interconnection constraints:}
The interconnection constraints between the primitive systems 1 and 2 are given by
\begin{equation}\label{interconnection_cond_simple} 
\dot{ \mathbf{q} }=\dot{\bar{ \mathbf{q} }}, \quad \mathbf{F} ^{\rm int}+\bar{ \mathbf{F} }^{\rm int}=0,
\end{equation} 
which indeed fits into the setting \eqref{interconnection_cond} with the distribution $ \Sigma _Q=\{( \dot{\mathbf{q}}, \dot{\bar{ \mathbf{q} }}, \dot  S) \mid \dot{ \mathbf{q} }=\dot{\bar{ \mathbf{q} }}\}$ for $Q= Q_1 \times Q_2 \times \mathbb{R} $. In this case, the interconnection constraints imply Newton's third law of action and reaction.

Juxtaposing the equations of the primitive systems 1 and 2, namely, equations \eqref{primitive_syst_1_simple} and \eqref{Prim2_SimpleSys} together with the interconnection constraints \eqref{interconnection_cond_simple} and eliminating the intermediate variables, the total system of equations is obtained as
\begin{equation}\label{interconnected_simple}
\left\{
\begin{array}{l}
\displaystyle\vspace{0.2cm}  \frac{d}{dt}\frac{\partial L_{\rm mech}}{\partial \dot{ \mathbf{q}}}- \frac{\partial L_{\rm mech}}{\partial  \mathbf{q} }=  \mathbf{F} ^{\rm fr}( \mathbf{q} , \dot{ \mathbf{q} }, S),
\\
\displaystyle -\frac{\partial U}{\partial S}\dot S=  \left\langle \mathbf{F} ^{\rm fr}(\mathbf{q},\dot{\mathbf{q}}, S), \dot{\mathbf{q}}\right\rangle .
\end{array} 
\right.
\end{equation}

\paragraph{Example: mass-spring-friction system.} A classical example is the mass-spring system with friction whose Lagrangian is $L(q,\dot{q},S)=L_{\rm mech}(q,\dot{q})-U(S)=\frac{1}{2}M\dot{q}^2-\frac{1}{2}Kq^2-U(S)$, with $M$ is the mass, $K$ is the spring constant, and $U(S)$ is the internal energy. By using \eqref{simple_systems}, in which the external force and the constraint $ \Delta _Q$ are absent, the equations of motion of the system are given by
\begin{equation}\label{MassSpringFrictionEqn}
M\ddot{q}+Kq=F^{\rm fr}(q,\dot{q}, S), \quad  -\frac{\partial U}{\partial S}\dot S=  F^{\rm fr}(q,\dot{q}, S)\dot{q},
\end{equation}
with the friction force $F^{\rm fr}(q,\dot{q}, S)=-R(q,S)\dot{q}$ for some friction coefficient $R>0$.

Juxtaposing the system equations of the primitive systems 1 and 2, namely, equations \eqref{primitive_syst_1_simple} and \eqref{Prim2_SimpleSys} together with the interconnection constraints, we get the system:
\begin{equation*}
\begin{split}
& M\ddot{q}+Kq= F^{\rm int}(q,\bar{q},\dot{q},\dot{\bar{q}}, S),\\
&\bar{F}^{\rm int}(q,\bar{q},\dot{q},\dot{\bar{q}}, S)+F^{\rm fr}(\bar{q},\dot{\bar{q}},S)=0,\quad 
\displaystyle -\frac{\partial U}{\partial S}\dot S=  F^{\rm fr}(\bar{q},\dot{\bar{q}}, S)\dot{\bar{q}},\\ 
&\dot{q}=\dot{\bar{q}}, \quad  F^{\rm int}(q,\bar{q},\dot{q},\dot{\bar{q}}, S)+ \bar{F}^{\rm int}(q,\bar{q},\dot{q},\dot{\bar{q}}, S)=0,
\end{split}
\end{equation*} 
from which the equations of motion \eqref{MassSpringFrictionEqn} are obtained by eliminating intermediate variables.

\color{black}

\subsection{Variational formulation for interconnected non-simple thermodynamic systems}\label{subsec_4c}

Let us consider the interconnection for the case of non-simple systems, which are thermodynamic systems with several entropies. We focus on the internal process of heat exchange.

\subsubsection{Thermodynamic systems with heat exchange}\label{heat_exchange}

\begin{figure}[h]
\vspace{-0.3cm}
\begin{center}
\includegraphics[scale=0.45]{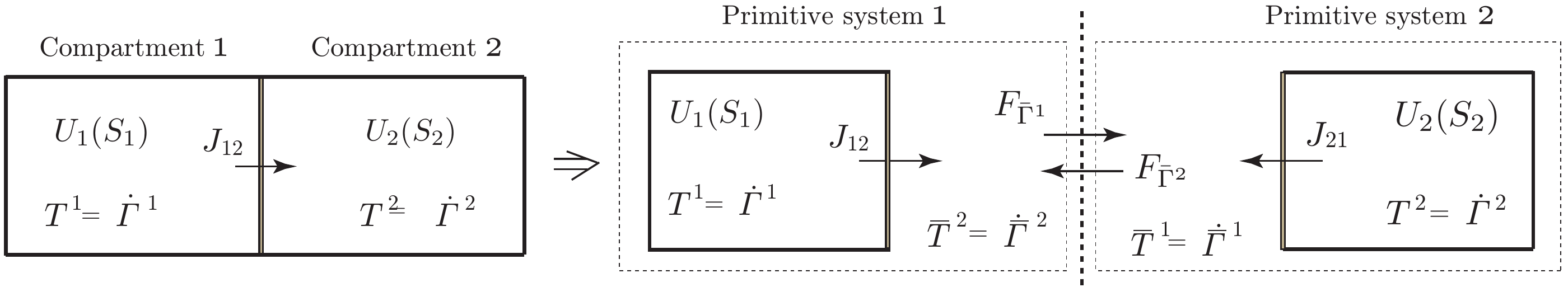}
\caption{Non-simple closed system experiencing heat conduction between two compartments}
\label{ClosedHeatTransSim}
\end{center}
%\vspace{-0.8cm}
\end{figure}

In order to illustrate the theory in simpler terms, we first develop the case of pure heat exchange, without any mechanical parts. We thus consider a non-simple adiabatically closed system experiencing heat conduction between two compartments, see Figure \ref{ClosedHeatTransSim}, where $S_k, T^k, \Gamma^k,\;k=1,2$ denote entropy, temperature, and thermal displacement of the $k$-th compartment. As earlier, we denote by $J_{k\ell}$, $k, \ell=1,2$, $k\neq \ell$ the entropy fluxes.

The Lagrangian for each primitive system reduces to the internal energy for each compartment, so that $L_k(S_k)= -U_k(S_k)$, $k=1,2$.
By tearing the original system into two primitive systems, we need to consider the \textit{intermediate thermodynamic forces}  $(F_{\bar{\Gamma}^1},F_{\bar{\Gamma}^2})$ dual to 
$(\dot{\bar{\Gamma}}^1,\dot{\bar{\Gamma}}^2)$ at the boundary of the two primitive systems.

\paragraph{Primitive system 1.} The variational condition for primitive system 1 is given as
\begin{equation*}
\delta \int_{t _1 }^{ t _2} \!\Big[ L_1\left(S_1\right)+  \dot{\Gamma }^1( S_1- \Sigma  _1)\Big] {\rm d}t +\int_{t _1 }^{ t _2} \left( F_{\Gamma^1} \delta{\Gamma^1}+F_{\bar{\Gamma}^2} \delta{\bar{\Gamma}^2} \right) {\rm d} t=0,
\end{equation*}
under the \textit{phenomenological constraint} and  \textit{variational constraint}
\begin{equation*}
\frac{\partial L_1}{\partial S_1}\dot \Sigma_1  =   J_{12}\dot{\bar{\Gamma}}^2, \qquad 
\frac{\partial L_1}{\partial S_1}\delta \Sigma_1  =  J_{12}\delta{\bar{\Gamma}}^2, 
\end{equation*}
with $ \delta \Gamma^1(t_k)=\delta \bar{\Gamma}^2(t_k)=0$, $k=1,2$.

Taking the variations we get
\begin{equation}\label{Primitive1_Conduction}
\displaystyle\vspace{0.2cm}\delta S_1:\;\dot{\Gamma}^{1}=- \frac{\partial L_{1}}{\partial S_{1}},
\qquad \delta  \Gamma ^1:\;\dot S_1= \dot{\Sigma}_{1}+F_{\Gamma^{1}}, 
\qquad 
\delta \bar \Gamma ^2:\;J_{12}+F_{\bar{\Gamma}^{2}}=0,
\end{equation}
while the phenomenological constraint gives
\begin{equation}\label{S_equ_primitive_1} 
\frac{\partial L_1}{\partial S_1} ( \dot  S_1 - F_{\Gamma^{1}})= J_{12} \dot {\bar{\Gamma }}^2.
\end{equation}

\paragraph{Primitive system 2.} Similarly, the variational condition for primitive system 2 is given as
\begin{equation*}
\delta \int_{t _1 }^{ t _2} \!\Big[ L_2\left(S_2\right)+  \dot{\Gamma }^2( S_2- \Sigma  _2)\Big] {\rm d}t +\int_{t _1 }^{ t _2} \left( F_{\Gamma^2} \delta{\Gamma^2}+F_{\bar{\Gamma}^1} \delta{\bar{\Gamma}^1} \right) {\rm d} t=0,
\end{equation*}
under the \textit{phenomenological constraint} and \textit{variational constraint}
\begin{equation*}
\frac{\partial L_2}{\partial S_2}\dot \Sigma_2  =  J_{21}\dot{\bar{\Gamma}}^1, \qquad 
\frac{\partial L_2}{\partial S_2}\delta \Sigma_2  =  J_{21}\delta{\bar{\Gamma}}^1, 
\end{equation*}
with $ \delta \Gamma^2(t_k)=\delta \bar{\Gamma}^1(t_k)=0$, $k=1,2$.

Taking the variations, we get
\begin{equation}\label{Primitive2_Conduction}
\displaystyle\vspace{0.2cm} \delta S_2:\;\dot{\Gamma}^{2}=- \frac{\partial L_{2}}{\partial S_{2}},
\qquad \delta \Gamma ^2:\; \dot S_2= \dot{\Sigma}_{2}+F_{\Gamma^{2}},
\qquad 
\delta \bar \Gamma ^1:\;J_{21}+F_{\bar{\Gamma}^{1}}=0,
\end{equation}
while the phenomenological constraint gives
\begin{equation}\label{S_equ_primitive_2} 
\frac{\partial L_2}{\partial S_2}( \dot  S_2 - F_{\Gamma^{2}})= J_{21} \dot {\bar{ \Gamma }}^1.
\end{equation}

\paragraph{Interconnection constraints.} The interconnection of the two systems is given by imposing the equality of the temperatures, thereby giving the interconnection constraint $\Sigma_{\rm heat}$ as
\begin{equation}\label{Int_HeatCond}
\Sigma_{\rm heat}:=\left\{(\dot{\Gamma}^1,\dot{\bar{\Gamma}}^2,\dot{\bar{\Gamma}}^1,\dot{\Gamma}^2) \;\middle|\; \dot{\Gamma}^{1}=\dot{\bar{\Gamma}}^{1}, \;\; \dot{\bar{\Gamma}}^{2}=\dot{\Gamma}^{2}  \right\}
\end{equation} 
from which its annihilator is obtained as
\begin{equation}\label{Int_HeatCond_annih}
\Sigma_{\rm heat}^{\circ}:=\left\{(F_{\Gamma^1},F_{\bar{\Gamma}^2},F_{\bar{\Gamma}^1},F_{\Gamma^2}) \;\middle|\; F_{\Gamma^1}+F_{\bar{\Gamma}^1}=0, \;\; F_{\bar{\Gamma}^2}+F_{\Gamma^2}=0 \right\} .
\end{equation}

Juxtaposing the equations of the primitive systems 1 and 2, namely, equations \eqref{Primitive1_Conduction}--\eqref{S_equ_primitive_2} together with the interconnection constraints \eqref{Int_HeatCond} and \eqref{Int_HeatCond_annih}  and eliminating the intermediate variables, yields the total system equations for heat exchange between two compartments; namely, one eventually gets the system
\[
T^1\dot S_1= J_{12}(T^1-T^2), \qquad T^2 \dot  S_2 = J_{12}(T^2-T^1),
\]
where we recall that $J_{12}=J_{21}$ is a priori assumed as before.
This approach also yields the variational formulation of the total system as an interconnected system of the two primitive systems, by adding the two action functionals, taking into account of both constraints as well as the variational constraint for interconnection $\delta{\Gamma}=(\delta{\Gamma}^1,\delta{\bar{\Gamma}}^2,\delta{\bar{\Gamma}}^1,\delta{\Gamma}^2)\in \Sigma_{\rm heat}$.

\subsubsection{Thermodynamic systems with mechanical interactions and heat exchange}

We consider now the case of an interconnected non-simple system with both mechanical interactions and heat exchange. 

The primitive systems that we consider are simple systems with a mechanical and a thermal part, also with possible mechanical and thermal interactions with their surrounding. Such systems themselves can be further decomposed as in \S\ref{subsec_4b}, however we consider each of them here as one primitive.

\begin{figure}[h]
\vspace{-0.3cm}\begin{center}
\includegraphics[scale=0.45]{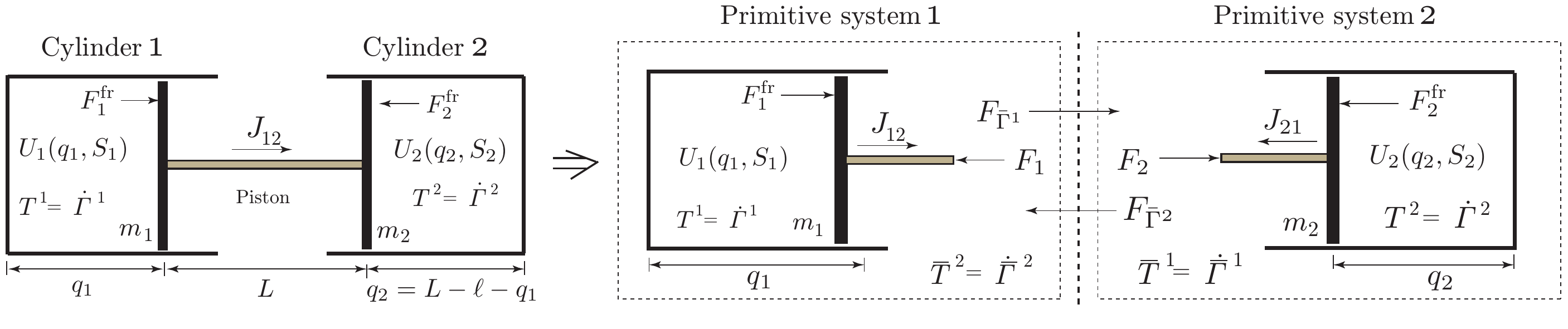}
\caption{Non-simple closed system with forcing and experiencing heat conduction between two compartments}
\label{ClosedHeatTransSim_piston}
\end{center}
\vspace{-0.6cm}
\end{figure}

For simplicity, we consider the case of only two subsystems ($P=2$) but we can easily generalize to the case of any number of subsystems in the context of interconnected systems. The Lagrangians of the primitive systems are $L_k:TQ_k \times \mathbb{R} \rightarrow \mathbb{R}$, $k=1,2$ and we suppose the possible occurrence of nonholonomic mechanical constraints $ \Delta _{Q_k} \subset TQ_k$, $k=1,2$.
A particular case of such a system is the celebrated piston problem, \cite{Gr1999}; see Figure \ref{ClosedHeatTransSim_piston}, which will be considered as an illustrative example later.

By combining the approach developed in the pure mechanical case in \S\ref{SubSubSec_VF_IntConnection} and for heat exchange in \S\ref{heat_exchange} we can state the following variational formulations.

\paragraph{Primitive system 1.} The variational formulation for the primitive system 1 is given as
\begin{equation*}
\delta \int_{t _1 }^{ t _2} \!\Big[ L_1\left( \mathbf{q} _1, \dot { \mathbf{q} }_1, S_1\right)+  \dot{\Gamma }^1( S_1- \Sigma  _1)\Big] {\rm d}t +\int_{t_1}^{t_2} \left\langle \mathbf{F} _{1}, \delta \mathbf{q} _1 \right\rangle {\rm d} t + \int_{t _1 }^{ t _2} \left( F_{\Gamma^1} \delta{\Gamma^1}+F_{\bar{\Gamma}^2} \delta{\bar{\Gamma}^2} \right) {\rm d} t=0,
\end{equation*}
subject to the \textit{kinematic constraint}
\begin{equation*}
\dot{ \mathbf{q} }_1 \in \Delta _{Q_1}( \mathbf{q} _1), \qquad \frac{\partial L_1}{\partial S_1}\dot \Sigma_1  =  \left\langle \mathbf{F} ^{\rm fr}_1, \dot{ \mathbf{q} }_1 \right\rangle + J_{12}\dot{\bar{\Gamma}}^2, 
\end{equation*}
for variations that are subject to the \textit{variational constraint}
\begin{equation*}
\delta  \mathbf{q} _1 \in \Delta _{Q_1}( \mathbf{q} _1), \qquad \frac{\partial L_1}{\partial S_1}\delta \Sigma_1  =  \left\langle \mathbf{F} ^{\rm fr}_1, \delta  \mathbf{q} _1 \right\rangle+ J_{12}\delta{\bar{\Gamma}}^2, 
\end{equation*}
with $ \delta \Gamma^1(t_k)=\delta \bar{\Gamma}^2(t_k)=0$, $k=1,2$.

Taking the variations, we get
\begin{equation}\label{Primitive1_Conduction_mech}
\begin{aligned} 
&\displaystyle\vspace{0.2cm}\delta S_1:\;\dot{\Gamma}^{1}=- \frac{\partial L_{1}}{\partial S_{1}},
\qquad \delta  \Gamma ^1:\;\dot S_1= \dot{\Sigma}_{1}+F_{\Gamma^{1}}, 
\qquad 
\delta \bar \Gamma ^2:\;J_{12}+F_{\bar{\Gamma}^{2}}=0,\\
&\displaystyle \delta \mathbf{q} _1:\;\;\frac{d}{dt} \frac{\partial L_1}{\partial \dot{\mathbf{q}}_1}- \frac{\partial L_1}{\partial \mathbf{q}_1} - \mathbf{F} _{1} - \mathbf{F} ^{\rm fr}_1\in \Delta_{Q_1}^\circ(\mathbf{q}_1),
\end{aligned} 
\end{equation}
while the kinematic constraints give
\begin{equation}\label{S_equ_primitive_1_mech} 
\dot{\mathbf{q}}_1 \in \Delta _{Q_1}( \mathbf{q}_1), \qquad \frac{\partial L_1}{\partial S_1} ( \dot  S_1 - F_{\Gamma^{1}})= \left\langle \mathbf{F} ^{\rm fr}_1, \dot{ \mathbf{q} }_1 \right\rangle + J_{12} \dot {\bar{\Gamma }}^2.
\end{equation}

\paragraph{Primitive system 2.} The variational formulation for the primitive system 2 is given as
\begin{equation*}
\delta \int_{t _1 }^{ t _2} \!\Big[ L_2\left( \mathbf{q} _2, \dot { \mathbf{q} }_2, S_2\right)+  \dot{\Gamma }^2( S_2- \Sigma  _2)\Big] {\rm d}t +\int_{t_1}^{t_2} \left\langle \mathbf{F} _{2},  \delta \mathbf{q} _2 \right\rangle {\rm d} t + \int_{t _1 }^{ t _2} \left( F_{\Gamma^2} \delta{\Gamma^2}+F_{\bar{\Gamma}^1} \delta{\bar{\Gamma}^1} \right) {\rm d} t=0,
\end{equation*}
subject to the \textit{kinematic constraint}
\begin{equation*}
\dot{ \mathbf{q} }_2 \in \Delta _{Q_2}( \mathbf{q} _2), \qquad \frac{\partial L_2}{\partial S_2}\dot \Sigma_2  =  \left\langle \mathbf{F} ^{\rm fr}_2, \dot{ \mathbf{q} }_2 \right\rangle + J_{21}\dot{\bar{\Gamma}}^1, 
\end{equation*}
for variations that are subject to the \textit{variational constraint}
\begin{equation*}
\delta  \mathbf{q} _2 \in \Delta _{Q_2}( \mathbf{q} _2), \qquad \frac{\partial L_2}{\partial S_2}\delta \Sigma_2  =  \left\langle \mathbf{F} ^{\rm fr}_2, \delta  \mathbf{q} _2 \right\rangle+ J_{21}\delta{\bar{\Gamma}}^1, 
\end{equation*}
with $ \delta \Gamma^2(t_k)=\delta \bar{\Gamma}^1(t_k)=0$, $k=1,2$.

Taking the variations, we get
\begin{equation}\label{Primitive2_Conduction_mech}
\begin{aligned} 
&\displaystyle\vspace{0.2cm}\delta S_2:\;\dot{\Gamma}^{2}=- \frac{\partial L_{2}}{\partial S_{2}},
\qquad \delta  \Gamma ^2:\;\dot S_2= \dot{\Sigma}_{2}+F_{\Gamma^{2}}, 
\qquad 
\delta \bar \Gamma ^1:\;J_{21}+F_{\bar{\Gamma}^{1}}=0,\\
&\displaystyle \delta \mathbf{q} _2:\;\;\frac{d}{dt} \frac{\partial L_2}{\partial \dot{\mathbf{q}}_2}- \frac{\partial L_2}{\partial \mathbf{q}_2} - \mathbf{F} _{2} - \mathbf{F} ^{\rm fr}_2\in \Delta_{Q_2}^\circ(\mathbf{q}_2),
\end{aligned} 
\end{equation}
while the kinematic constraints give
\begin{equation}\label{S_equ_primitive_2_mech} 
\dot{\mathbf{q}}_2 \in \Delta _{Q_2}( \mathbf{q}_2), \qquad \frac{\partial L_2}{\partial S_2} ( \dot  S_2 - F_{\Gamma^{2}})= \left\langle \mathbf{F} ^{\rm fr}_2, \dot{ \mathbf{q} }_2 \right\rangle + J_{21} \dot {\bar{\Gamma }}^1.
\end{equation} 

\paragraph{Interconnection constraints.} The interconnection of the two systems is given by considering the distribution
\begin{equation}\label{Int_HeatCond_mech} 
\Sigma _{\rm int}= \Sigma _{Q} \times \Sigma_{\rm heat},
\end{equation} 
where $\Sigma_{\rm heat}$ is associated to the equality of temperatures as in \eqref{Int_HeatCond} and $ \Sigma _Q \subset T(Q_1 \times Q_2)$ describes the mechanical interaction as in \S\ref{SubSubSec_VF_IntConnection}.
From this the annihilator is obtained as
\begin{equation}\label{Int_HeatCond_mech_annih}
\!\!\Sigma_{\rm int}^{\circ}:=\left\{( \mathbf{F} _{1}, \mathbf{F} _{2},F_{\Gamma^1},F_{\bar{\Gamma}^2},F_{\bar{\Gamma}^1},F_{\Gamma^2}) \;\middle|\; (\mathbf{F} _1, \mathbf{F} _2) \in \Sigma _Q^\circ,\, F_{\Gamma^1}+F_{\bar{\Gamma}^1}=0, \, F_{\bar{\Gamma}^2}+F_{\Gamma^2}=0 \right\} .
\end{equation}

Juxtaposing the equations of the primitive systems 1 and 2, namely, equations \eqref{Primitive1_Conduction_mech}--\eqref{S_equ_primitive_2_mech} together with the interconnection constraints \eqref{Int_HeatCond_mech} and \eqref{Int_HeatCond_mech_annih}  and eliminating the intermediate variables, yields the total system equations for the non-simple thermodynamic system, as derived in \S\ref{subsub_2ci}, see in particular equations \eqref{nonsimple_systems} and their rewriting in \eqref{nonsimple_systems_rewriting} (for $P=2$).

This approach also directly yields the variational formulation of the total system, by adding the two action functionals, taking into account of both constraints as well as the variational constraint for interconnection $( \delta \mathbf{q} , \delta{\Gamma})=( \delta \mathbf{q} _1, \delta \mathbf{q} _2,\delta{\Gamma}^1,\delta{\bar{\Gamma}}^2,\delta{\bar{\Gamma}}^1,\delta{\Gamma}^2)\in ( \Delta _{Q_1} \times \Delta _{Q_2} \times \mathbb{R} ^4) \cap \Sigma_{\rm int}$.

\begin{figure}[h]
\vspace{-0.3cm}\begin{center}
\includegraphics[scale=0.45]{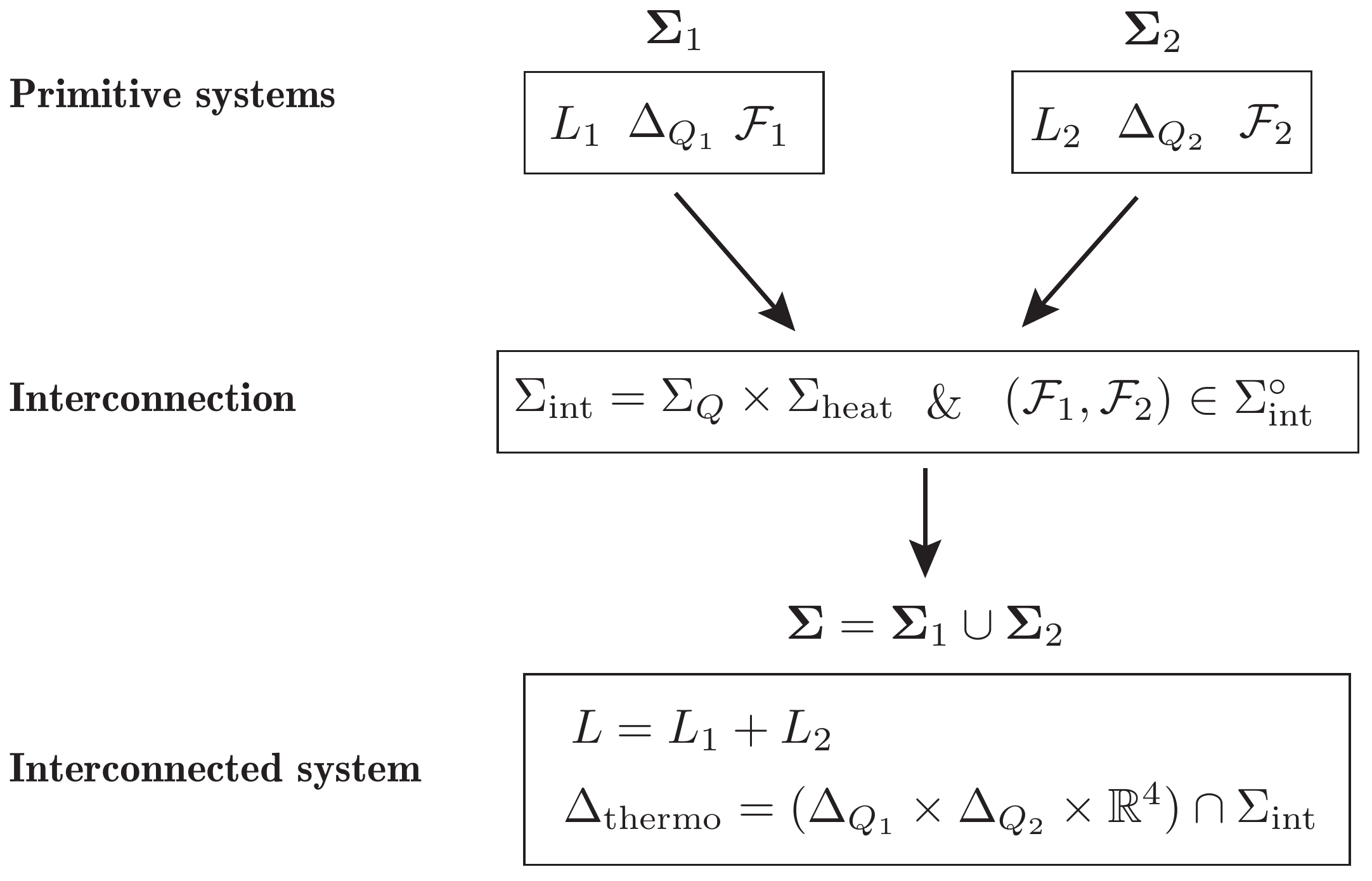}
\caption{Interconnection in thermodynamics}
\label{InterconnectionThermodynamics}
\end{center}
\vspace{-1.2cm}
\end{figure}

The Figure \ref{InterconnectionThermodynamics} illustrates how the primitive systems $ \boldsymbol{\Sigma} _1$ and $ \boldsymbol{\Sigma} _2$ are interconnected to yield the original system $ \boldsymbol{\Sigma} $, where $ \mathcal{F} _1=( \mathbf{F} _1, F_{ \Gamma _1}, F_{\bar{ \Gamma }_2})$ and $ \mathcal{F} _2=( \mathbf{F} _2, F_{ \Gamma _2}, F_{\bar{ \Gamma }_1})$ are the interconnection forces.

\paragraph{Example: the piston problem.} As illustrated in
Figure \ref{ClosedHeatTransSim_piston}, consider the piston-cylinder system that is composed of two cylinders connected by a movable piston and suppose that each cylinder contains an ideal gas. Suppose also that the system is isolated. Note that the dynamics of this system is known as the so-called adiabatic piston problem when the piston is adiabatic, and there has been some controversy about the final equilibrium state of this system; see \cite{Gr1999} as to the history of this problem.

Now we set $Q_1=Q_2= \mathbb{R} $ and the Lagrangian and friction force of $k$-th subsystem is
\[
L_k(q_k, \dot  q_k, S_k)= \frac{1}{2} m_k \dot  q_k ^2 - U_k(q_k, S_k) , \quad F^{\rm fr}_k(q_k, \dot  q_k, S_k)= - \lambda ^k (q_k, S_k) \dot  q _k, \quad k=1,2.
\]
The mechanical interconnection constraint is $ \Sigma _Q=\{ ( \dot  q_1, \dot  q_2) \mid \dot{q}_2=-\dot{q}_1\}$, which follows from the holonomic constraint $q_2=L-\ell-q_1$, see Figure \ref{ClosedHeatTransSim_piston}, hence its annihilator is $ \Sigma _Q^\circ=\{(F_1, F_2) \mid F_1= F_2\}$. The interconnection $\Sigma_{\rm heat}$ is associated to the equality of temperatures as in \eqref{Int_HeatCond}. From this, and given $J_{12}= - \kappa $, the conduction coefficient, we can explicitly write the equations \eqref{Primitive1_Conduction_mech}--\eqref{S_equ_primitive_2_mech} for each subsystems, as well as the interconnection conditions \eqref{Int_HeatCond_mech_annih}. It is easily seen to give the system of differential equations
\[
\left\{
\begin{array}{l}
\vspace{0.1cm}(m_1+m_1) \ddot q = p_1A_1-p_2 A_2- (\lambda _1+ \lambda _2) \dot  q,\\
\vspace{0.1cm}T^1 \dot  S_1=  \kappa (T^2-T^1)+ \lambda _1 \dot  q ^2,\\
 T^2 \dot  S_2=  \kappa (T^1-T^2)+ \lambda _2 \dot  q ^2 ,
\end{array}
\right.
\]
where we define $q:= q^1$, which completely describe the evolution of the system. The adiabatic case corresponds to $ \kappa =0$.

\section{Conclusions}

In this paper we have reviewed a variational formulation for nonequilibrium thermodynamics that extends the Hamilton principle of classical mechanics to include irreversible processes, by focusing on the process of friction and heat exchange. We have also proposed a new variational formulation of interconnected systems, where the notion of interconnections in network theory is exclusively extended to nonequilibrium thermodynamics. First we have illustrated that the variational formulation is of the Lagrange-d'Alembert type, in the sense that it consists of a critical curve condition subject to two kind of constraints: a kinematic (phenomenological) constraint and a variational constraint. It is based on the notion of thermodynamic displacement associated to each irreversible process. Second, we have presented a modeling approach to multiphysical complicated thermodynamic systems, in which we first reticulate the system to identify the underlying primitive systems with their interconnection in clear mathematical terms. Further, we have developed such an approach based on the variational formulations, by constructing a variational setting for primitive systems as well as the interconnection conditions in the form of a distribution and its annihilator, $ \Sigma _{\rm int}$ and $ \Sigma _{\rm int}^\circ$, valid for both mechanical and temperature conditions. This setting allows one to concatenate the variational principles in such a way to recover the variational principle of the overall interconnected thermodynamic system. We have illustrated the setting with elementary examples, while we have left the treatment of other process, such as diffusion, chemical reactions, as well as continuum systems, for a future work.

\color{black} 
\vskip6pt

\paragraph{Acknowledgement.} H.Y. is partially supported by JSPS (Grant-in-Aid for Scientific Research 22K03443), JST CREST(JPMJCR1914), Waseda University (SR 2023C-089), and the MEXT ``Top Global University Project'' at Waseda University.

%%%%%%%%%% Insert bibliography here %%%%%%%%%%%%%%

%%%

%%
\end{document}